\documentclass[10pt,journal,compsoc]{IEEEtran}  

\IEEEoverridecommandlockouts                              

\usepackage{graphicx} 
\usepackage{epsfig} 
\usepackage{algorithmic}
\usepackage{array}
\usepackage{times} 
\usepackage{amsmath} 
\usepackage{amssymb}  
\usepackage{graphicx}
\usepackage{epstopdf}
\usepackage{algorithmic}
\usepackage{bm}
 \usepackage{dblfloatfix}
 \usepackage{url}
 \usepackage{color}
\ifCLASSOPTIONcompsoc
  \usepackage[caption=false,font=footnotesize,labelfont=sf,textfont=sf]{subfig}
\else
  \usepackage[caption=false,font=footnotesize]{subfig}
\fi
\ifCLASSOPTIONcompsoc
\usepackage[nocompress]{cite}
\else
\usepackage{cite}
\fi
\newtheorem{Theorem 1}{Theorem}
\newtheorem{Theorem 2}[Theorem 1]{Theorem}
\newtheorem{Theorem 3}[Theorem 1]{Theorem}
\newtheorem{Theorem 4}[Theorem 1]{Theorem}
\newtheorem{Definition 1}{Definition}
\newtheorem{Definition 2}[Definition 1]{Definition}
\newtheorem{Remark 1}{Remark}
\newtheorem{Remark 2}[Remark 1]{Remark}
\newtheorem{Remark 3}[Remark 1]{Remark}
\newtheorem{Lemma 1}{Lemma}
\newtheorem{Lemma 2}[Lemma 1]{Lemma}
\newtheorem{Lemma 3}[Lemma 1]{Lemma}
\newtheorem{Lemma 4}[Lemma 1]{Lemma}

\date{}
\begin{document}
\title{
Distributed  Event Localization via Alternating Direction Method of
Multipliers }

\author{Chunlei Zhang and Yongqiang Wang
    \thanks{*The work was supported in part by the Institute for Collaborative Biotechnologies through grant W911NF-09-0001.}
    \thanks{Chunlei Zhang and Yongqiang Wang are with the
        department of Electrical and Computer Engineering, Clemson
        University, Clemson, SC 29634, USA
        {\tt\small \{chunlez,yongqiw\}@clemson.edu}}%
}

\thispagestyle{empty}
\pagestyle{empty}

\IEEEtitleabstractindextext{
	\begin{abstract}	
		This paper addresses the problem of distributed event localization
		using noisy range measurements with respect to sensors with known
		positions. Event localization is fundamental in many wireless sensor
		network applications  such  as homeland  security, law enforcement,
		and environmental studies. However, most existing distributed algorithms require the target event to be within the convex hull of the deployed sensors.  Based on the alternating direction method of multipliers
		(ADMM), we propose 	two scalable distributed algorithms named GS-ADMM and J-ADMM
		which do not require the target event to be within the convex hull of the deployed sensors. More specifically, the two algorithms can be implemented in a scenario in which the
		entire sensor network is divided into several clusters with cluster heads collecting measurements within each cluster and exchanging intermediate computation information to achieve localization consistency (consensus)	across all clusters. This scenario is important in many applications such as homeland security and law enforcement. Simulation
		results confirm effectiveness of the proposed algorithms. 
	\end{abstract}	
	
	}

\maketitle
\thispagestyle{empty}
\IEEEpeerreviewmaketitle
\IEEEraisesectionheading{\section{Introduction}\label{sec:introduction}}
\IEEEPARstart{W}{ith} the ability to transmit/receive information and fuse data,
smart sensors  enabled and greatly advanced numerous applications
such as environmental monitoring \cite{bonnet2000querying}, target
tracking \cite{cevher2008distributed}, underwater detection
\cite{heidemann2005underwater}, and acoustic gunfire localization
\cite{george2013shooter}, \cite{simon2004sensor}. Among these
applications, \emph{event localization} is a significant and essential
component or even the ultimate goal. Taking the gunfire localization as an
example, if some threat sources or impulsive events (e.g., shooting
or explosion) occur, it is of imperative importance to localize
these threat sources to make prompt reactions (e.g., giving warning,
providing aid). In fact, sensor network based event localization has received significant attentions and plenty of techniques have been proposed in the literature, using either
angle-of-arrival measurements
\cite{kaplan2001maximum,schmidt1986multiple,gavish1992performance}, time-of-arrival (ToA) (including
time-difference-of-arrival, i.e., TDoA) measurements
\cite{ho2004accurate,yang2009efficient}, or received signal strength (RSS) \cite{blatt2006energy,shi2008distributed,wang2009sensor,meng2011diffusion,wang2011semidefinite,wang2013new,yuan2015exact,zhang2015distributed}. There are also some work that discussed the event localization problem based on noisy range measurements directly, which can be obtained based on ToA, TDoA, or RSS information \cite{beck2008exact,jia2011set,pinar2010convex,fu2012complex,ouguz2014angular}. Generally speaking, these existing methods for event localization formulate the localization problem as a maximum likelihood estimation problem \cite{pinar2010convex} or a least squares problem \cite{yuan2015exact}, which is solved by minimizing the non-convex objective function iteratively \cite{blatt2006energy} or by applying various convex relaxations \cite{yang2009efficient}. 

From the implementation point of view, existing event localization algorithms can be cast into two categories:  centralized approaches and distributed approaches. Centralized approaches always gather (noisy) measurements (e.g.,
range measurements) obtained by all sensors to a
processing center, which then estimates the event location using a
certain centralized optimization algorithm. Typical centralized methods include the parallel projection method \cite{jia2011set}, convex relaxation plus semidefinite programming (SDP) or second-order cone programming method \cite{yang2009efficient,pinar2010convex,wang2011semidefinite,fu2012complex,wang2013new,ouguz2014angular,yuan2015exact}.  However, a severe shortcoming of centralized localization
algorithms is that the computation complexity at the processing
center might be quite high which poses great challenges for low-cost
sensor nodes with limited computational capabilities. In addition,
the required communication to collect all measurements to a
single central node may be problematic due to possible traffic
bottleneck and severe  constraints on communication ranges.
Moreover, once the central node fails due to, e.g., attacks or power
depletion, the entire network slips into a state of paralysis.
Therefore, techniques solving the event localization problem in a
distributed way are crucial for sensor network based event
localization.

In contrast to centralized algorithms, distributed localization
algorithms are designed to run the computation over the entire
network instead of on a  processing center. In general, distributed algorithms
are often established on massive parallelism or sequential
calculations and mutual collaboration
\cite{bachrach2005localization}. So compared with centralized
algorithms, distributed designs have better scalability,
flexibility, and failure resilience. One typical distributed approach for event localization is projection-based algorithms which solve the event localization problem by projecting an initial estimate onto sensing disks \cite{blatt2006energy}, circles \cite{shi2008distributed,wang2009sensor,wang2015decentralized}, or rings \cite{zhang2015distributed}. However, these projection-based localization algorithms are very sensitive to the initial values when the target event lies outside the convex hull of sensors, as will be shown in Sec. 6.

This paper is motivated by acoustic event localization which is crucial on battlefields \cite{deligeorges2015mobile}. In such applications,  the target event has no communication or computation
capability,  which differentiates the problem from sensor localization problems in which the locations of sensors are estimated \cite{simonetto2014distributed}. Furthermore, in such applications, the target events lie outside the convex hull of deployed sensors, which renders existing projection-based algorithms inappropriate. SDP relaxation based algorithms can avoid the convex hull problem and are traditionally employed to solve the event localization problem \cite{yang2009efficient,pinar2010convex,wang2011semidefinite,fu2012complex,wang2013new,ouguz2014angular,yuan2015exact}. However, as far as we known, existing SDP relaxation based algorithms for event localization are all centralized, with a central node collecting and processing all data, which makes them susceptible to  processing center failure and traffic bottleneck. In this paper, we propose two distributed event localization approaches based on a clustered architecture motivated by mobile acoustic localization applications such as the PinPoint\textsuperscript{TM}  system from BioMimetics Systems Inc.  The PinPoint\textsuperscript{TM} mobile localization sensor network can be deployed as a mobile infrastructure for impulsive
threat event detection and localization
\cite{deligeorges2015mobile,cakiades2012fusion}. Each  PinPoint\textsuperscript{TM} sensor is a small omnidirectional microphone array which localizes impulsive acoustic events by correlating the ToA measurements among its microphone  cells.  
 In fact, since each sensor 
has an integrated microphone array, individual sensors are able to identify and localize
a target event without assistance or cooperation with other sensors. However, due
to close distances between the microphone cells, the accuracy of individual sensors is very limited and unsatisfactory, and collaboration among the
sensors is necessary to improve localization accuracy \cite{deligeorges2015mobile,cakiades2012fusion}.

The above application motivated us to assume a localization
architecture in which an entire network is divided into several
clusters. A cluster head (which can be a regular sensor) collects
and fuses measurements (e.g., noisy ranges) obtained from all members in its cluster. Two cluster heads in
different clusters can exchange information (the local estimates of target events) if a communication link
is available between them; otherwise they don't have access to each
other's information.  Our developed algorithms can also be applied in some other applications where a cluster-based architecture is employed. A typical example is the wide-area
monitoring and control in large-scale power systems
\cite{nabavi2015distributed}, \cite{giannakis2013monitoring}. To estimate the electro-mechanical
oscillation modes, a large number of phasor measurement units (PMU)
have to be deployed across a power network to conduct measurements. The
measurements from  PMUs have to be fused to diagnose the
inter-area oscillation modes. However, wide-area communication between PMUs is
very expensive \cite{hasan2009analyzing}.   To fuse information
across the PMUs without imposing heavy communication overhead, a
similar structure as ours is adopted in
\cite{nabavi2015distributed}, \cite{giannakis2013monitoring}. Other examples on cluster-based architecture can be found in \cite{venkatesan2007coordinating,papadogiannis2008dynamic,zhang2009networked,jain2016backhaul}.

The core of our distributed localization algorithms is  the
alternating direction method of multipliers (ADMM), which has been
proven extremely suitable in distributed convex optimization,
especially for large-scale problems  \cite{boyd2011distributed}. The key idea of ADMM
is to obtain a global solution through the cooperation of small
local subproblems. ADMM is easy to parallelize and implement,
and is robust to noise and computation errors \cite{simonetto2014distributed}. Our proposed
localization approaches take full advantages of ADMM
which enables local optimizations within individual clusters as
subproblems. Then through cooperation of subproblems in neighboring
clusters,  a global event localization could be reached. That is to
say, the estimated locations obtained by individual clusters are
made as consistent as possible. Such consistency is of crucial
importance in many applications. For example, when a sporadic impulsive event
requiring immediate responsive actions is detected by several
monitors, consistency in the estimated location across 
monitors is the key for multiple monitors to coordinate cooperative
operations.

\emph{Contribution:} The main contribution of this paper is two
ADMM-based distributed event localization algorithms, i.e., GS-ADMM and J-ADMM. Compared with existing centralized  SDP relaxation based algorithms for event localization, the two algorithms divide the computation on a central node to different clusters to avoid possible center failure and traffic bottleneck, and in the mean time, guarantee consistency of the estimates across all clusters among which only limited communications are available. Furthermore, the two algorithms take advantages of SDP relaxation to avoid the convex hull problem compared with existing projection-based algorithms. Moreover,
the algorithms are proven to converge with a convergence rate of
$O(1/t)$ where $t$ is the iteration time.

\emph{Organization:}  The rest of this paper is organized as
follows: Section  2 states the
formulation of the problem. To solve the problem, a convex
relaxation is required and the method proposed by
\cite{simonetto2014distributed} is recapitulated in Section
3. In Section 4, two algorithms named
GS-ADMM and J-ADMM  are proposed based on ADMM, with  their
convergence properties analyzed in Section
 5. Section
 6 gives numerical simulation
results. In the end, a conclusion is made in Section
 7.

\section{PROBLEM STATEMENT}
\begin{figure}[!t]
   \centering
        \includegraphics[width=8cm]{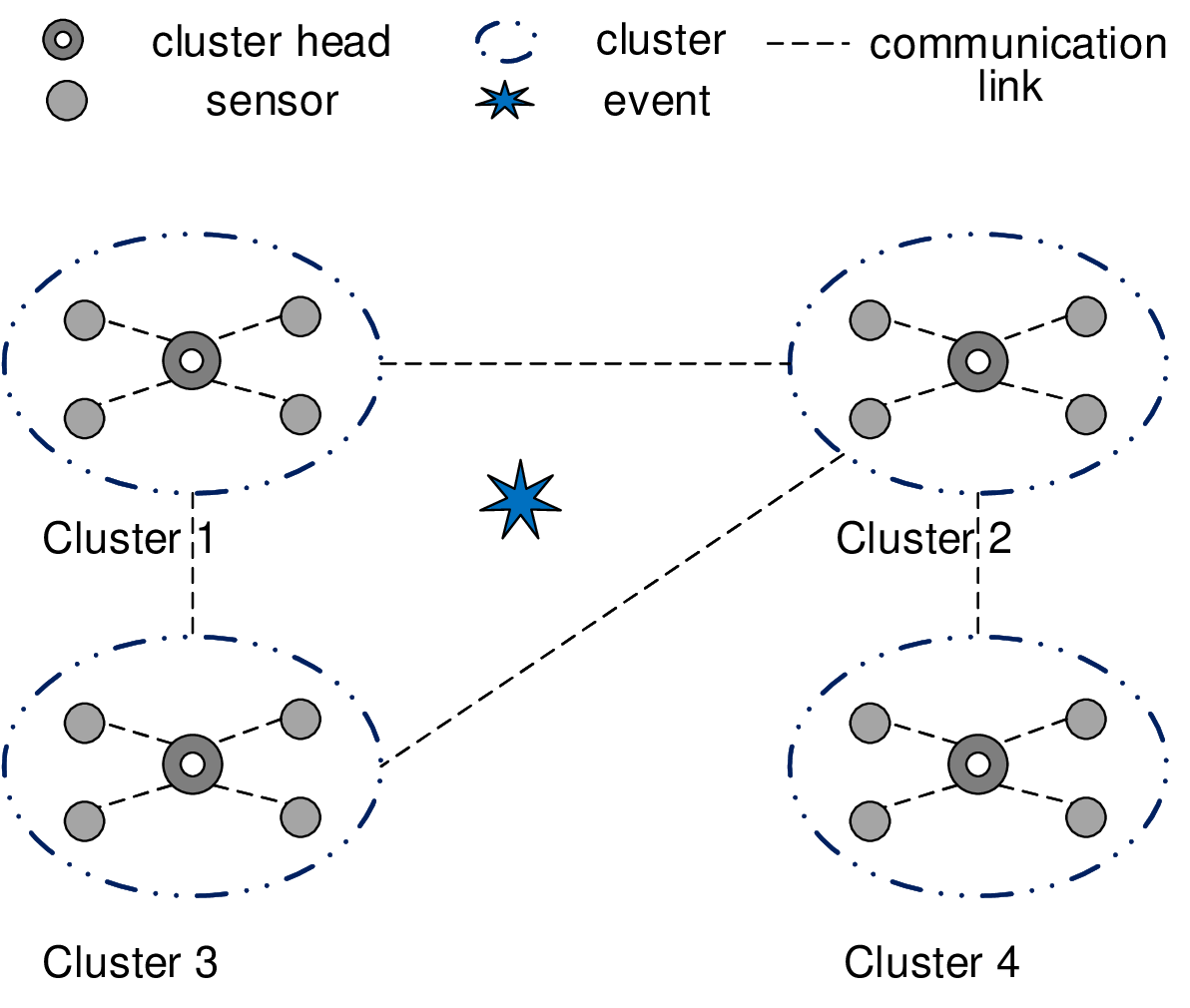} 
    \caption{Cluster based event localization architecture ($m=4$)}
    \label{fig:event localization structure illustration}
\end{figure}

Motivated by mobile acoustic event localization applications such as the PinPoint\textsuperscript{TM} event localization sensor network \cite{deligeorges2015mobile,cakiades2012fusion}, we consider  a  localization sensor network divided into  $m$
  clusters (cf. Fig. \ref{fig:event localization structure illustration} for the case $m=4$).
 Denote the number of constituent sensors of cluster $i$ as ${N}_{i}$
 ($i=1,2,\ldots,m$). We consider localization in $D$ ($D\in\{1,2,3\}$) dimensional
 Euclidean space
 and suppose that the position of the target event is denoted as
 $\bm{x}\in\mathbb{R}^{D}$.
 Denote the position of the $k$th sensor in the $i$th cluster as $\bm{a}_{i,k}\in\mathbb{R}^{D}$.
 The $k$th sensor in the $i$th cluster can obtain a noisy
 range measurement ${r}_{i,k}$ of its distance with respect to a target event:
\begin{eqnarray}\label{eq:measurement model}
{r}_{i,k}={d}_{i,k}+{v}_{i,k} \nonumber
\end{eqnarray}
where  ${d}_{i,k}=\parallel \bm{x}-\bm{a}_{i,k} \parallel$ denotes the actual
distance between the event position and the $k$th sensor of the
$i$th cluster, and ${v}_{i,k}$ is the Gaussian noise term.

Then the event localization problem amounts to estimating the
unknown event location $\bm{x}$ using known sensor positions $\bm{a}_{i,k}$
and noisy range measurements ${r}_{i,k}$
($i=1,2,\ldots,m,\,k=1,2,\ldots,N_i$). Still motivated by acoustic event localization applications (e.g., the  PinPoint\textsuperscript{TM} event localization sensor network \cite{deligeorges2015mobile,cakiades2012fusion}), we assume that a cluster head exists in each
cluster $i$, which can gather range measurements ${r}_{i,k}$
from all sensors within the cluster. In addition, a cluster head can
communicate and exchange information with the cluster head of a
neighboring cluster if there is a communication link between them
(cf. Fig. \ref{fig:event localization structure illustration}). In
this case, we also say that these two clusters can communicate.  We
assume that the communication pattern forms a
 connected network, i.e., there is a (multi-hop) path (composed of multiple communication links connected in succession) between any pair of cluster heads.
For example, in Fig. \ref{fig:event localization structure illustration}, cluster 1 is able to exchange
information with clusters 2 and 3 (via cluster heads); cluster 2
can exchange information with clusters 1, 3, and 4 (via cluster
heads), etc. Denote
 ${B}_{i}$ as the set of all neighboring clusters of cluster $i$, 
$\hat{B}_i$ as the union of set ${B}_{i}$ and cluster $i$ itself, and $|B_i|$ as the number of clusters in ${B}_{i}$.

As in most existing results, we use the maximum likelihood
method for event localization
\cite{pinar2010convex,fu2012complex}.  Let
${p}_{i,k}({d}_{i,k}(\bm{x},\bm{a}_{i,k})|{r}_{i,k})$ denote the measuring
probability density function (PDF) for sensor $k$ in cluster $i$ and assume that it is a log-concave function of unknown distance ${d}_{i,k}$ \cite{simonetto2014distributed},  we can
write this problem using the maximum likelihood method (which is costly but efficient \cite{beck2008iterative}):
\begin{eqnarray}\label{eq:ML formulation}
{\bm{x}}_{\textrm{ML}}^*={\textrm{argmax}}_{\bm{x}\in\mathbb{R}^{D}}\sum\limits_{i=1}^{m}\sum\limits_{k=1}^{{N}_{i}}\textrm{ln}{p}_{i,k}({d}_{i,k}(\bm{x},\bm{a}_{i,k})|{r}_{i,k}).
\end{eqnarray}

\section{CONVEX RELAXATION}
Problem \eqref{eq:ML formulation} is non-convex and it is generally infeasible to
find a global optimal solution \cite{simonetto2014distributed}. So a convex relaxation is needed to convert problem \eqref{eq:ML formulation} into a
convex optimization problem. Following the idea of \cite{simonetto2014distributed}, we use an SDP based relaxation approach.  However, it is worth noting that there are inherent differences between the problem considered here and the sensor-position estimation problem in \cite{simonetto2014distributed} where each sensor with unknown position estimates its own position using embedded computation capability. The differences are evident from the following example. Suppose that there is only one target to localize. In the case of \cite{simonetto2014distributed}, the target will be a sensor with unknown position and it estimates its own position \emph{alone} using a \emph{centralized} SDP based on all information gathered from adjacent sensors, including their positions and corresponding range measurements. Whereas in our case, the target is an event without any communication or computation capability and the event position estimation process is conducted \emph{cooperatively} in a \emph{distributed} way among the clusters.

To facilitate the  relaxation, we first define the following new
variables: $y=\bm{x}^T\bm{x}$, ${\epsilon}_{i,k}={d}_{i,k}^2$. Then we stack
${\epsilon}_{i,k}, k\in\{1,2,...,N_i\}$ into $\bm{\epsilon}_{i}$ and further stack $\bm{\epsilon}_{i},i\in\{1,2,...,m\}$ into
$\bm{\epsilon}\triangleq[\bm{\epsilon}_1^T,
\bm{\epsilon}_2^T,...,\bm{\epsilon}_m^T]^T$. In the same way we stack
${d}_{i,k}$  into $\bm{d}_{i}$ and
$\bm{d}\triangleq[\bm{d}_{1}^T,\bm{d}_{2}^T,...,\bm{d}_{m}^T]^T$. Then the
cost function can be written as
\begin{eqnarray}
f(\bm{d}) =- \sum\limits_{i=1}^{m}\sum\limits_{k=1}^{{N}_{i}}\textrm{ln}{p}_{i,k}({d}_{i,k}|{r}_{i,k}). \nonumber
\end{eqnarray}

Consider the case of white zero-mean Gaussian noise, i.e.,
${v}_{i,k}\sim\mathcal{N}(0,{\sigma}_{i,k}^2)$, then the above problem
 can be rewritten as
\begin{equation}\label{eq:cost function of guassian noise}
\begin{aligned}
f(\bm{d}) =\sum\limits_{i=1}^{m}\sum\limits_{k=1}^{{N}_{i}}{\sigma}_{i,k}^{-2}({d}_{i,k}^2-2{d}_{i,k}{r}_{i,k}+{r}_{i,k}^2) 
    \end{aligned}
\end{equation}

Without loss of generality, we can set the standard deviation $\sigma_{i,k}$ in \eqref{eq:cost function of guassian noise} to one. Now, problem \eqref{eq:ML formulation} can be relaxed into the
following constrained optimization problem:
\begin{equation}\label{eq:constrained optimization nonconvex set}
\begin{aligned}
\mathop {\min }\limits_{\bm{x},\bm{\epsilon},\bm{d},y} \qquad & f(\bm{d}) \\
\textrm{subject to }\qquad & y-2\bm{x}^T\bm{a}_{i,k}+\parallel\bm{a}_{i,k}\parallel^2={\epsilon}_{i,k}, \quad y=\bm{x}^T\bm{x},   \\
&{\epsilon}_{i,k}={d}_{i,k}^2, {d}_{i,k}\ge 0,   \\
&\forall i\in\{1,2,...,m\},\quad k\in\{1,2,...,{N}_{i}\}.
\end{aligned}
\end{equation}

However, in this case, the constraints of \eqref{eq:constrained optimization nonconvex set} still define a non-convex set \cite{simonetto2014distributed}. Using  Schur
complements \cite{ouellette1981schur}, the following convex
relaxation can be obtained:
\begin{equation} \label{eq:constrained optimization convex set}
\begin{aligned}
\mathop {\min }\limits_{\bm{x},\bm{\epsilon},\bm{d},y}  \qquad & f(\bm{d})  \\
\textrm{subject to}  \qquad &  y-2\bm{x}^T\bm{a}_{i,k}+\parallel\bm{a}_{i,k}\parallel^2={\epsilon}_{i,k}, \quad {\epsilon}_{i,k}\ge 0 , \\
&\begin{pmatrix}
1 & {d}_{i,k} \\
{d}_{i,k} & {\epsilon}_{i,k}
\end{pmatrix}\succeq 0,\quad {d}_{i,k}\ge 0,  \\
&\forall i\in\{1,2,...,m\}, \quad k\in\{1,2,...,{N}_{i}\},  \\
&  \begin{pmatrix}
\bm{I_D} & \bm{x} \\
\bm{x}^T & y
\end{pmatrix}\succeq 0, \quad y\ge 0.
\end{aligned}
\end{equation}

Problem \eqref{eq:constrained optimization convex set} is a convex problem with inequality constraints
\cite{boyd2011distributed}. We can rewrite the cost function as
\begin{equation} \label{eq:constrained optimization convex set 2}
\begin{aligned}
f(\bm{d,\epsilon})= \sum\limits_{i=1}^{m}\sum\limits_{k=1}^{{N}_{i}}{\sigma}_{i,k}^{-2}({\epsilon}_{i,k}-2{d}_{i,k}{r}_{i,k}+{r}_{i,k}^2)
\end{aligned}
\end{equation}
 by enforcing a change of variables ${\epsilon}_{i,k}=d_{i,k}^2$ to further relax it to a semidefinite programming (SDP)
problem \cite{simonetto2014distributed}. Now, we can propose ADMM based solutions for problem \eqref{eq:constrained optimization convex set}.

\section{PROPOSED DISTRIBUTED ALGORITHMS}
\subsection{Preliminaries: Standard ADMM}
ADMM is an algorithm which is suitable to solve problems in the
following form \cite{boyd2011distributed}:
\begin{equation}\label{eq:admm}
\begin{aligned}
\min_{\bm{x},\bm{z}} \qquad & f(\bm{x}) + g(\bm{z}) \\
\textrm{subject to}  \qquad & A\bm{x} + K\bm{z} = \bm{c}.
\end{aligned}
\end{equation}
where $\bm{x}\in\mathbb{R}^{n}$, $\bm{z}\in\mathbb{R}^{m}$,
$A\in\mathbb{R}^{p\times n}$, $K\in\mathbb{R}^{p\times m}$, and
$\bm{c}\in\mathbb{R}^{p}$, and   $f(\bm{x})$ and $g(\bm{z})$ are convex functions.
To get the optimal value $p^* = \textrm{inf}\{f(\bm{x})+g(\bm{z})\mid
A\bm{x}+K\bm{z}=\bm{c}\}$ for problem \eqref{eq:admm}, one can first form an augmented Lagrangian function:
\begin{eqnarray}
\lefteqn{ \mathcal{L}_{\rho}(\bm{x},\bm{z},\bm{\mu}) = f(\bm{x})+g(\bm{z}) {} }\nonumber \\
& &  \quad\quad\quad+\bm{\mu}^T(A\bm{x}+K\bm{z}-\bm{c})+\frac{\rho}{2} \parallel A\bm{x}+K\bm{z}-\bm{c}\parallel^{2}, \nonumber
\end{eqnarray}
where $\bm{\mu}$ is the Lagrange multiplier associated with the constraint
$A\bm{x}+K\bm{z}=\bm{c}$ and $\rho>0$ is a predefined penalty parameter.  Then ADMM
solves problem (\ref{eq:admm}) by updating $\bm{x},\bm{z},\bm{\mu}$ in the following
sequence: first an $\bm{x}$-minimization step \eqref{eq:x
update}, then a $\bm{z}$-minimization step \eqref{eq:z update},
and finally a dual variable update \eqref{eq:y update}:
\begin{eqnarray}
&& \bm{x}^{k+1}=\textrm{argmin}_{\bm{x}} \mathcal{L}_{\rho}(\bm{x},\bm{z}^k,\bm{\mu}^k), \label{eq:x update} \\
&& \bm{z}^{k+1}=\textrm{argmin}_{\bm{z}} \mathcal{L}_{\rho}(\bm{x}^{k+1},\bm{z},\bm{\mu}^k), \label{eq:z update}\\
&& \bm{\mu}^{k+1}=\bm{\mu}^k+\rho(A\bm{x}^{k+1}+K\bm{z}^{k+1}-\bm{c}). \label{eq:y update}
\end{eqnarray}

Next, we will propose two distributed algorithms for
event localization using the framework of standard ADMM.
\subsection{Problem Reformulation}
In distributed algorithms, neighboring
nodes have to generate and exchange copies of local estimates
to ensure a consistent global estimation across all nodes. In
our event localization architecture, a cluster is treated as a normal
node which solves a common event localization problem based
on measurements obtained by sensors within the cluster. And
neighboring clusters exchange intermediate computational results
(through cluster heads) to guarantee that all clusters reach the
same estimation value. 

To better interpret our algorithms, we define a local vector
\begin{eqnarray}
\quad\bm{p}_{i}\triangleq(\bm{\epsilon}_{i}^T,\bm{d}_{i}^T,y_i,\bm{x}_i^T)^T\in\mathbb{R}^{2N_i+D+1},\quad
i\in\{1,2,...,m\}, \nonumber
\end{eqnarray}
which is owned by cluster $i$.

We let $\bm{p}$ denote the stacked
vector of $\bm{p}_{i}$ and define a convex set
\begin{eqnarray}
\mathcal{P}_{i} \triangleq\{\bm{p}_{i}|\bm{p}_{i} \quad \textrm{verifies} \quad \eqref{eq:constrained optimization convex set}\}. \nonumber
\end{eqnarray}

Then problem \eqref{eq:constrained optimization convex set} can be rewritten as
\begin{equation}
\begin{aligned}
\mathop {\min }\limits_{\bm{p}} \qquad & f(\bm{p}) \\
\textrm{subject to} \qquad &  \bm{p}_{i}\in  \mathcal{P}_{i}, \quad \forall i\in\{1,2,...,m\}, \label{eq:admm model}
\end{aligned}
\end{equation}
where, in our situation, $f(\bm{p})$ is given as follows:
\begin{eqnarray}
f(\bm{p})=-
\sum\limits_{i=1}^{m}\sum\limits_{k=1}^{{N}_{i}}\ln{p}_{i,k}({d}_{i,k}|{r}_{i,k})=\sum\limits_{i=1}^{m}{f}_i(\bm{p}_{i}).
\label{eq: objective function for admm model}
 \end{eqnarray}

\subsection{ADMM based problem formulation}
From the architecture in \eqref{eq: objective function for admm model}, it is easy to see that problem
\eqref{eq:admm model} can be divided into $m$ subproblems, which can be solved in a distributed
 way using ADMM by adding some constraints on $\bm{p}_i$. Next we present
the basic idea based on a graph-based formulation of the
communication pattern.

Using graph theory \cite{bondy1976graph}, the communication pattern
of cluster heads can be represented by $G=\{V,E\}$, where the
set $V$ denotes the set of cluster heads, and $E$ denotes the set of
undirected edges (communication links) between clusters. We use
$e_{i,j}\in E, i<j$ to denote the link (if there is) between cluster
heads $i$ and $j$. We use $|E|$ to represent the total number of
undirected edges. In our problem formulation, each cluster is
associated with a local cost function $f_i(\bm{p}_i)$, and all
clusters work together to solve the problem in \eqref{eq:admm model}. Assume that the local cost function $f_i$
is only known to cluster $i$, then to reach consistency (consensus)
of estimated position values among all clusters, we impose a
constraint $\bm{x}_i=\bm{x}_j$ if there exists an edge $e_{i,j}\in E$
between clusters $i$ and $j$. Introduce a matrix $J_i=[0_{D\times(2N_i+1)},I_D]\in\mathbb{R}^{D\times(2N_i+D+1)}$, where $I_D$ denotes the $D$ dimensional identity matrix, then $\bm{x}_i$ can be represented as $\bm{x}_i=J_i\bm{p}_i$. So the constraint $\bm{x}_i=\bm{x}_j$ can be represented as $J_i\bm{p}_i=J_j\bm{p}_j$. 

Now we are able to rewrite problem \eqref{eq:admm model} into a distributed ADMM
form as follows:
\begin{equation}\label{eq:distributed admm form}
\begin{aligned}
\mathop {\min }\limits_{\bm{p}_{i},\,i\in\{1,2,...,m\}} \qquad & \sum\limits_{i=1}^{m}f_i(\bm{p}_{i})  \\
\textrm{subject to} \qquad & J_i\bm{p}_{i}=J_j\bm{p}_{j},\quad \forall e_{i,j}\in E,  \\
& \bm{p}_{i}\in \mathcal{P}_{i}, \quad \forall i\in\{1,2,...,m\},
\end{aligned}
\end{equation}
or in a more compact way:
\begin{equation}
\begin{aligned}
\mathop {\min }\limits_{\bm{p}} \qquad& f(\bm{p}) \label{eq:compact distributed admm form} \\
\textrm{subject to} \qquad&  CJ\bm{p}=0, \quad \bm{p}_{i}\in  \mathcal{P}_{i}, \quad \forall i\in\{1,2,...,m\},
\end{aligned}
\end{equation}
where $\bm{p}=[\bm{p}_1^T, \bm{p}_2^T,...,\bm{p}_m^T]^T$, $J=\textrm{diag}\{J_1,J_2,\dots,J_m\}\in\mathbb{R}^{mD\times(\sum\limits_{i=1}^{m}2N_i+D+1)}$, and $C$ is the
edge-node incidence matrix of graph $G$ as defined in
\cite{wei2012distributed}. For example, in the one-dimensional case
($D=1$), $C=[c_{i,j}]$ is an $|E|\times m$ matrix whose $|E|$
rows correspond to the $|E|$ edges and $m$ columns correspond to
the $m$ clusters such that:
\begin{eqnarray}
c_{i,j}=\left\{\begin{matrix}
1& \textrm {if the } i^{th} \textrm{ edge originates at cluster }  j,  \\
-1& \textrm {if the } i^{th} \textrm{ edge terminates at cluster }  j, \\
0& \textrm{otherwise}.
\end{matrix}\right.
\end{eqnarray}
Here we define that each edge $e_{i,j}$ originates at $i$ and
terminates at $j$.

It can be easily verified that the incidence matrix $C$ for Fig.
\ref{fig:event localization structure illustration}
 is
\begin{eqnarray}\label{eq:C}
C=\begin{bmatrix}
1&  -1& 0 & 0 \\
0& 1  & -1 & 0 \\
1 & 0 & -1 & 0\\
0&  1& 0 & -1
\end{bmatrix}.
\end{eqnarray}

For high dimensional cases, where $D\ge2$, $C\in
\mathbb{R}^{|E|D\times mD}$ can be obtained by replacing the value of $1$
and $-1$ with $I_D$ and $-I_D$, respectively, with $I_D$ denoting the $D$
dimensional identity matrix. Then the $C$ matrix for Fig. 1 becomes
\begin{eqnarray}
C=\begin{bmatrix}
I_D&  -I_D& 0_D & 0_D \\
0_D& I_D  & -I_D & 0_D \\
I_D & 0_D & -I_D & 0_D\\
0_D&  I_D& 0_D & -I_D
\end{bmatrix}.
\end{eqnarray}

In this formulation, after each cluster obtains its local estimate
$\bm{p}_i$, it sends the value $J_i\bm{p}_i$ (estimated event position $\bm{x}_i$) to neighboring clusters. By adding
the constraint $J_i\bm{p}_{i}=J_j\bm{p}_{j}, \forall i\in\{1,2,...,m\},
j\in{B}_{i} \nonumber $ as shown in \eqref{eq:distributed admm form}, the consistency of
individual  event position $J_i\bm{p}_i$ ($\bm{x}_i$) estimated across the clusters is
guaranteed. Now we are in place to present our detailed algorithms to
solve \eqref{eq:distributed admm form}.

\begin{Remark 2} \label{Remark:3}
      Note that although a normal way to apply ADMM to consensus problems is to create auxiliary local variables (cf. \cite{simonetto2014distributed}), we just put the constraint $J_i\bm{p}_i=J_j\bm{p}_j$ directly here. The reason that we omit the auxiliary local variables is to save storage space at each cluster, since auxiliary local variables take additional storage space.  Furthermore, by adding the constraint $J_i\bm{p}_i=J_j\bm{p}_j$, we can have both a sequential and a parallel realization with convergence guaranteed, which will be detailed in the following subsection. This kind of constraint and its induced ADMM algorithm is called extended ADMM, which is discussed and applied in many recent work, e.g., \cite{mota2013d,nabavi2015distributed,deng2017parallel,wei2012distributed,yuan2015communication}.   
\end{Remark 2}

\subsection{Proposed Algorithms}
Let $\bm{\lambda}_{i,j}$ be the Lagrange multiplier relevant to the constraint $J_i\bm{p}_{i}=J_j\bm{p}_{j}$. Then the regularized augmented Lagrangian function of problem \eqref{eq:distributed admm form} can be reformulated as
\begin{equation}\label{eq:augemented lagrangian function}
\begin{aligned}
&\mathcal{L}_{\rho}(\bm{p},\bm{\lambda})=\sum\limits_{i=1}^{m}f_i(\bm{p}_{i}) 
\\
&\qquad+ \sum\limits_{e_{i,j}\in E}(\bm{\lambda}_{i,j}^T(J_i\bm{p}_i-J_j\bm{p}_{j})+\frac{\rho}{2}\parallel J_i\bm{p}_i-J_j\bm{p}_{j} \parallel ^2),
\end{aligned}
\end{equation}
where $\bm{\lambda}_{i,j}$ are stacked into $\bm{\lambda}_{i}$ for all $j\in {B}_i$ and $\bm{\lambda}_{i}$ are stacked into $\bm{\lambda}$ for all $i\in\{1,2,...,m\}$. 

Applying ADMM, we can get the following two updating recursions:
\begin{eqnarray}
&&\bm{p}^{t+1}=\textrm{argmin}_{\bm{p}_i\in \mathcal{P}_i}\mathcal{L}_{\rho}(\bm{p},\bm{\lambda}^{t}), \label{eq:pi update}\\
&&\bm{\lambda}_{i,j}^{t+1}=\bm{\lambda}_{i,j}^{t}+\rho(J_i\bm{p}_i^{t+1}-J_j\bm{p}_j^{t+1}). \label{eq:lambda update}
\end{eqnarray}
Here, we can update $\bm{p}$ in two different ways.  One way is based on the
Gauss-Seidel update \cite{golub2012matrix} in which clusters update
in a sequential order. The other way is the Jacobian scheme in which
all clusters update in parallel \cite{saad2003iterative}.

\textbf{Gauss-Seidel update (GS-ADMM):} We first consider an algorithm based on the
Gauss-Seidel update. Gauss-Seidel update for distributed ADMM has
been explored theoretically and proven able to converge in most
cases for convex objective functions (see, e.g., \cite{he2012alternating,hong2017linear,tao2011recovering}).  GS-ADMM based solution for
distributed event localization can be described as follows:\\
\begin{flushleft}
    \vspace{-1.22cm}
    \rule{0.49\textwidth}{0.2pt}
\end{flushleft}
\vspace{-0.2cm}

\textbf{Algorithm \uppercase\expandafter{\romannumeral1}: GS-ADMM }
\begin{flushleft}
    \vspace{-0.83cm}
    \rule{0.49\textwidth}{0.2pt}
\end{flushleft}

 Each cluster initializes $\bm{p}_i^{0}$, $\bm{\lambda}_{i,j}^{0}$.

 \textbf{Input:} $\bm{p}_i^{t}$, $\bm{\lambda}_{i,j}^{t}$

 \textbf{Output:} $\bm{p}_i^{t+1}$, $\bm{\lambda}_{i,j}^{t+1}$

\begin{enumerate}

\item All clusters update  their local vectors in a sequential order and send their local vectors $J_i\bm{p}_i^{t+1}$ to neighboring clusters in $B_i$ immediately, where \label{item:1}
\begin{small}
\begin{equation}\label{eq:GS-ADMM pi update}
\begin{aligned}
\lefteqn{\bm{p}_i^{t+1}=\textrm{argmin}_{\bm{p}_i\in \mathcal{P}}f_i(\bm{p}_{i})+} \\
&\sum\limits_{j\in\hat{B}_i,j\ge i}(\bm{\lambda}_{i,j}^{tT}(J_i\bm{p}_{i}-J_j\bm{p}_j^{t})+ \frac{\rho}{2}\parallel J_i\bm{p}_{i}-J_j\bm{p}_{j}^{t} \parallel ^2)+ \\
&\sum\limits_{j\in\hat{B}_i, j<i}(\bm{\lambda}_{i,j}^{tT}(J_i\bm{p}_{i}-J_j\bm{p}_j^{t+1})+ \frac{\rho}{2}\parallel J_i\bm{p}_{i}-J_j\bm{p}_{j}^{t+1} \parallel ^2). 
\end{aligned}
\end{equation}
\end{small}
 Here we also consider the effect of $J_i\bm{p}_i^{t}$ when updating $\bm{p}_i^{t+1}$ by adding a term $\frac{\rho}{2}\parallel J_i\bm{p}_{i}-J_i\bm{p}_{i}^{t}\parallel^2$. Problem (\ref{eq:GS-ADMM pi update}) with $f_i$ given in \eqref{eq:constrained optimization convex set 2} is an SDP problem that can be solved by common convex toolboxes such as Yalmip
\cite{simonetto2014distributed,lofberg2004yalmip}, which is used in
our simulations. 

\item Each cluster computes
\begin{eqnarray}
\bm{\lambda}_{i,j}^{t+1}=\bm{\lambda}_{i,j}^{t}+\rho(J_i\bm{p}_{i}^{t+1}-J_j\bm{p}_j^{t+1}). \label{eq:GS-ADMM lambda update}
\end{eqnarray}

\item Set $t=t+1$, and go to 1).

\end{enumerate}

\begin{flushleft}
    \vspace{-0.8cm}
    \rule{0.49\textwidth}{0.2pt}
\end{flushleft}

 In GS-ADMM, all clusters update their local estimated position values in  a sequential way just as some projection-based algorithms. Sequential update can be used in small-size networks. For large-scale networks, a parallel method is more appropriate. So we also propose another algorithm based on Jacobian scheme which is amendable for parallelization.

 \textbf{Jacobian based ADMM (J-ADMM)}: Algorithm J-ADMM is motivated by the work in \cite{deng2017parallel}, which proposed the Proximal Jacobian ADMM by  adding some proximal terms when updating $\bm{p}_i$.
 We adopt the same idea here and prove that if the proximal  terms meet some additional requirements, convergence of this algorithm can be guaranteed.
  The detailed procedure of  J-ADMM is given as follows, with the convergence analysis detailed in the following section.\\
\begin{flushleft}
    \vspace{-1.2cm}
    \rule{0.49\textwidth}{0.2pt}
\end{flushleft}
\vspace{-0.2cm}

\textbf{Algorithm \uppercase\expandafter{\romannumeral2}: J-ADMM }
\begin{flushleft}
    \vspace{-0.83cm}
    \rule{0.49\textwidth}{0.2pt}
\end{flushleft}

Each cluster initializes $\bm{p}_i^{0}$, $\bm{\lambda}_{i,j}^{0}$.

\textbf{Input:} $\bm{p}_i^{t}$, $\bm{\lambda}_{i,j}^{t}$

\textbf{Output:} $\bm{p}_i^{t+1}$, $\bm{\lambda}_{i,j}^{t+1}$

\begin{enumerate}

\item Each cluster updates its local vector in parallel:
\begin{equation}\label{eq:J-ADMM pi update}
\begin{aligned}
\lefteqn{\bm{p}_i^{t+1}=\textrm{argmin}_{\bm{p}_i\in \mathcal{P}}f_i(\bm{p}_{i})\quad } \\
&+ \sum\limits_{j\in\hat{B}_{i}}(\bm{\lambda}_{i,j}^{tT}(J_i\bm{p}_{i}-J_j\bm{p}_j^{t})+ \frac{\rho}{2}\parallel J_i\bm{p}_{i}-J_j\bm{p}_{j}^{t} \parallel ^2) \\
&+ \frac{\rho\gamma_i}{2}\parallel J_i\bm{p}_i-J_i\bm{p}_i^{t} \parallel^2 .
\end{aligned}
\end{equation}

The last term of the above equality, i.e., $\frac{\rho\gamma_i}{2}\parallel
J_i\bm{p}_i-J_i\bm{p}_i^{t} \parallel^2$, is  the proximal term we
added where $\gamma_i\ge 0$ is a scalar.  Problem (\ref{eq:J-ADMM pi update}) with $f_i$ given in \eqref{eq:constrained optimization convex set 2} is an SDP problem that can be solved
by common convex toolboxes such as Yalmip
\cite{simonetto2014distributed,lofberg2004yalmip},  which is used in
our simulations.

\item Each cluster sends its local vector $J_i\bm{p}_i^{t+1}$ to neighboring clusters in $B_i$.

\item Each cluster computes
\begin{eqnarray}
\bm{\lambda}_{i,j}^{t+1}=\bm{\lambda}_{i,j}^{t}+\rho(J_i\bm{p}_{i}^{t+1}-J_j\bm{p}_j^{t+1}). \label{eq:J-ADMM lambda update}
\end{eqnarray}

\item Set $t=t+1$, and go to 1).
\end{enumerate}
\begin{flushleft}
    \vspace{-0.8cm}
    \rule{0.49\textwidth}{0.2pt}
\end{flushleft}

\begin{Remark 1}\label{Remark:1}
    A distinct  difference between GS-ADMM and J-ADMM
    is  the way they update $\bm{p}_i$. In GS-ADMM, each cluster updates
    its local estimated position value in a sequential way, which
    requires a globally predefined order. Whereas in J-ADMM, all clusters
    update  their local estimated position values simultaneously.
    We remark that GS-ADMM is appropriate for small-scale sensor networks. But for
    large-scale networks, updating in a sequential way may be quite
    time-consuming and parallel methods like J-ADMM are more appropriate. So different updating methods should be
    chosen according to the size of networks and other practical concerns.
\end{Remark 1}

 In fact, if we disregard the PinPoint\textsuperscript{TM} motivated application scenario, the proposed two algorithms can be completely distributed to each sensor by allowing sensors to have access to neighboring sensors' positions and range measurements with respect to the target event. However, we argue that this, in fact, may cost more energy since each sensor has to solve an SDP problem.  In addition, the required storage  overhead is larger since each sensor has to store neighboring sensors' positions and range measurements. Furthermore, consider a situation where two sensors can communicate with each other and have the same neighbors. Then the position estimation process conducted at these two sensors are the same, which leads to redundant processing of the same data. While in our clustered architecture, only cluster heads need to conduct position estimation and in fact, each sensor in the cluster can take turns to be the cluster head, which is helpful to average energy consumption. Compared with the iterative schemes, e.g., projection-based algorithms, where each sensor only has access to its own position and range measurement, our algorithms are \emph{insensitive} to the convex hull problem. And compared with centralized SDP-based algorithms, our clustered architecture is robust to processing center failure or traffic bottleneck problems. In addition, the convex relaxation methods used at each cluster can be further improved by using recent works such as \cite{yang2009efficient,pinar2010convex,wang2011semidefinite,fu2012complex,wang2013new,ouguz2014angular,yuan2015exact}. 

\section{CONVERGENCE ANALYSIS}
In this section, we analyze the convergence properties of GS-ADMM
and J-ADMM. As our algorithms are applications of distributed ADMM,
the analysis benefits from many existing results on general
distributed ADMM \cite{wei2012distributed,han2012note,mota2013d}.

\subsection{Convergence Analysis of GS-ADMM}
 Let  
 $\bm{p}^k=[\bm{p}_1^{kT},\bm{p}_2^{kT},...,\bm{p}_m^{kT} ]^T$ and $\bm{\lambda}^k=[\bm{\lambda}_{i,j}^k]_{ij,e_{i,j}\in E}$ be the
  iterates generated by algorithm GS-ADMM following \eqref{eq:GS-ADMM pi update} and \eqref{eq:GS-ADMM lambda update}. Assume that
  the initial problem \eqref{eq:distributed admm form} admits a solution $(\bm{p}^*,\bm{\lambda}^*)$, i.e.,  the
  Lagrangian function $L(\bm{p},\bm{\lambda})=f(\bm{p})+\bm{\lambda}^TCJ\bm{p}$ has a saddle point (note: not the augmented Lagrangian function),  then the following theorem
  holds:
\begin{Theorem 1}\label{Theorem:GS-ADMM Convergence} Let
$\bar{\bm{p}}^{t+1}=\frac{1}{t+1}\sum\limits_{k=0}^{t}\bm{p}^{k+1}$
be the average of $\bm{p}^{k}$ up  to iteration time $t+1$, then the followings hold for all $t$:

(1)
\begin{equation}\label{eq:item 1}
0\le
L(\bar{\bm{p}}^{t+1},\bm{\lambda}^*)-L(\bm{p}^*,\bm{\lambda}^*)\le
\frac{c_0}{t+1},
\end{equation}

(2) The sequence $(\bm{p}_1^k,\bm{p}_2^k,...,\bm{p}_m^k)$ deduced
by GS-ADMM  converge to $(\bm{p}_1^*,\bm{p}_2^*,...,\bm{p}_m^*)$, i.e., $\lim\limits_{k\to\infty}\parallel \bm{p}^k-\bm{p}^*\parallel=0$. In addition, we have
$J_1\bm{p}_1^*=J_2\bm{p}_2^*=...=J_m\bm{p}_m^*$.  

Here 
\begin{equation}
\begin{aligned}
 \lefteqn{c_0=\frac{1}{2\rho}\parallel \bm{\lambda}^0-\bm{\lambda}^*\parallel^2} \\
 & \quad +\frac{\rho}{2}(\parallel HJ(\bm{p}^0-\bm{p}^*)\parallel^2+\parallel J\bm{p}^0-J\bm{p}^*\parallel^2),\label{eq:c0}
 \end{aligned}
\end{equation}
and $H=\min\{0,C\}$ ($H_{i,j}=\min\{0,C_{i,j}\}$).
 \end{Theorem 1}

{\it Proof}:	(\ref{eq:item 1}) can be obtained following a way similar to Theorem 4.4
	in \cite{wei2012distributed}. A detailed proof is given in
	Appendix A. 
	To prove the second statement, recall that the objective function is
		\begin{eqnarray}
		f(\bm{d}) =\sum\limits_{i=1}^{m}\sum\limits_{k=1}^{{N}_{i}}{\sigma}_{i,k}^{-2}({d}_{i,k}^2-2{d}_{i,k}{r}_{i,k}+{r}_{i,k}^2). \nonumber
		\end{eqnarray}
		Setting
		$h_{i,k}={\sigma}_{i,k}^{-2}({d}_{i,k}^2-2{d}_{i,k}{r}_{i,k}+{r}_{i,k}^2)$,
		we have $f(\bm{d})
		=\sum\limits_{i=1}^{m}\sum\limits_{k=1}^{{N}_{i}}h_{i,k}$. Note
		that
		$h_{i,k}$  is a quadratic function and is
		strongly convex. Since the sum of strongly convex functions is still
		strongly convex,  our objective function $f(\bm{d})$ is strongly
		convex. Further note that $f(\bm{p})$ is equal to $f(\bm{d})$ and the set $\mathcal{P}_{i}$ is convex and closed. Therefore, our problem satisfies the requirements of both strongly convex objective function and convex-and-closed constraint set in \cite{han2012note}. Now we proceed to prove the second statement. First, rewriting $CJ\bm{P}=0$ in the form of $\sum\limits_{i=1}^{m}[C]_iJ_i\bm{p}_i=0$, where $[C]_i$ denotes the columns of $C$ associated with cluster $i$, we can form a variational inequality $MVI(Q,U)$ similar to (5)-(6) in \cite{han2012note}:
		$$\langle\bm{u}-\bm{u}^*,\bm{Q}(\bm{u}^*)\rangle\ge0, \quad \forall \bm{u}\in \mathcal{U},$$
		where
		\begin{displaymath}
		\begin{aligned}
		&{ \bm{u}^*:=\left(\begin{array}{c}
			\bm{p}_1^*\\
			\bm{p}_2^*\\
			\cdots\\
			\bm{p}_m^*\\
			\bm{\lambda}^*
			\end{array}\right),}
		\quad
		{ \bm{Q}(\bm{u}^*):=\left(\begin{array}{c}
			\xi_1^*+J_1^T[C]_1^T\bm{\lambda}^*\\
			\xi_2^*+J_2^T[C]_2^T\bm{\lambda}^*\\
			\cdots\\
			\xi_m^*+J_m^T[C]_m^T\bm{\lambda}^*\\
			CJ\bm{p}
			\end{array}\right),}\\
		&{\mathcal{U}:=\prod\limits_{i=1}^{m}\mathcal{P}_i\times \mathbb{R}^{|E|D}}.
		\end{aligned}
		\end{displaymath}
		Then following the proof of Lemma 4.1 in \cite{han2012note}, we can get that $(\bm{p}_1^{k+1},...,\bm{p}_m^{k+1},\bm{\lambda}^{k+1})$ is a solution to  $MVI(Q,U)$  if $CJ\bm{p}=0$ and $[C]_iJ_i\bm{p}_i^{k}=[C]_iJ_i\bm{p}_i^{k+1}$ hold. Secondly, following the proof of Lemma 4.2 in \cite{han2012note}, we can get the following inequality:
		\begin{small}
			\begin{eqnarray}\nonumber
			&&\langle\bm{\lambda}^*-\bm{\lambda}^k,CJ\bm{p}\rangle\ge\sum\limits_{i=1}^{m}\omega_i\parallel\bm{p}_i^{k+1}-\bm{p}_i^*\parallel^2+\rho\parallel CJ\bm{p}^{k+1}\parallel^2\\ \nonumber
			&&+\rho\sum\limits_{i=1}^{m}\langle [C]_iJ_i\bm{p}_i^{k+1}-[C]_iJ_i\bm{p}_i^{*},\sum\limits_{j=i+1}^{m}([C]_jJ_j\bm{p}_j^{k}-[C]_jJ_j\bm{p}_j^{k+1})\rangle \\
			&&-\rho\sum\limits_{i=1}^{m}\langle [C]_iJ_i\bm{p}_i^{k+1}-[C]_iJ_i\bm{p}_i^{*},\frac{1}{|B_i|}([C]_iJ_i\bm{p}_i^{k+1}-[C]_iJ_i\bm{p}_i^{k})\rangle, \nonumber
			\end{eqnarray}
		\end{small}
		where $f_i(\bm{p}_i)$ is strongly convex with modulus $\omega_i$. Thirdly, define an auxiliary block-diagonal matrix $M$:
		\begin{small}
			\begin{displaymath}
			M=\left(\begin{array}{cccc}
			\rho mJ_1^T[C]_1^T[C]_1J_1 &  \ldots & 0 & 0\\
			\cdots &  \ddots & \cdots & \cdots\\
			0 &  \ldots & \rho mJ_m^T[C]_m^T[C]_mJ_m & 0\\
			0 &  \ldots & 0 & \rho^{-1}I
			\end{array}\right).
			\end{displaymath}
		\end{small}
		Then by following the idea of the proof of Lemma 4.3 in \cite{han2012note}, the following inequality can be obtained:
		\begin{eqnarray}\nonumber
		&&\parallel \bm{u}^{k+1}-\bm{u}^{*} \parallel_M^2\le \parallel \bm{u}^{k}-\bm{u}^{*} \parallel_M^2 \\ 
		&&\qquad\qquad-2\sum\limits_{i=1}^{m}\omega_i\parallel\bm{p}_i^{k+1}-\bm{p}_i^*\parallel^2-\rho\parallel CJ\bm{p}^{k+1}\parallel^2 \nonumber \\ 
		&&\qquad\qquad+3m\rho\sum\limits_{i=1}^{m}\parallel [C]_iJ_i\bm{p}_i^{k+1}- [C]_iJ_i\bm{p}_i^{*}\parallel^2, \nonumber
		\end{eqnarray}
		where 
		\begin{equation}\nonumber
		\begin{aligned}
		&\parallel \bm{u}\parallel_M^2:= \parallel\bm{\lambda}\parallel_{\rho^{-1}}^2+ \\
		&\rho m(\parallel [C]_1J_1\bm{p}_1\parallel^2+\parallel [C]_2J_2\bm{p}_2\parallel^2+...+\parallel [C]_mJ_m\bm{p}_m\parallel^2). \nonumber
		\end{aligned}
		\end{equation}
		Finally, when $0<\rho<\min \limits_{1\le i \le m}\{\frac{2\omega_i}{3m\parallel [C]_iJ_i\parallel^2}\}$ holds, we can get the second statement following the proof of Theorem 4.1 in \cite{han2012note}.

\hfill $\blacksquare$

\begin{Remark 2} \label{Remark:2}
     Recall $\bm{\lambda}^{k+1}=\bm{\lambda}^{k}+\rho C J\bm{p}^{k+1}$,  we can get
     \begin{eqnarray}
     \lefteqn{\bm{\lambda}^{k+1}=\bm{\lambda}^{k}+\rho C J\bm{p}^{k+1} }\nonumber \\
     &&=\bm{\lambda}^{k-1}+\rho CJ (\bm{p}^{k+1}+\bm{p}^{k})=...=\bm{\lambda}^{0}+\rho CJ \sum\limits_{i=1}^{k+1}\bm{p}^{i}.  \nonumber
     \end{eqnarray}
     When $k\to \infty$, we have $ \bm{\lambda}^{k+1}\to \bm{\lambda}^{*}$. In other words, $\bm{\lambda}^{*}=\bm{\lambda}^{0}+\rho CJ \sum\limits_{i=1}^{\infty}\bm{p}^{i}$. So $c_0$ can be  represented as:
     \begin{eqnarray}
    \lefteqn{c_0=\frac{\rho}{2}\parallel CJ \sum\limits_{i=1}^{\infty}\bm{p}^{i}\parallel^2} \nonumber\\
    &&+\frac{\rho}{2}(\parallel HJ(\bm{p}^0-\bm{p}^*)\parallel^2+\parallel J\bm{p}^0-J\bm{p}^*\parallel^2). \nonumber
     \end{eqnarray}
     
    It is clear that $c_0$ will increase with an increase in $\rho$, so if the iteration time $t$ is fixed, $L(\bar{\bm{p}}^{t+1},\bm{\lambda}^*)-L(\bm{p}^*,\bm{\lambda}^*)$
    will also increase with an increase in $\rho$. That is to say, with $\rho$
    increasing, the iteration time to reach convergence will
    increase, namely convergence rate will be slower. Although with an increase in $\rho$, the convergence rate will
    decrease, $\rho$ cannot be too small. This is because if $\rho$ is
    too small, the constraint  $J_i\bm{p}_i=J_j\bm{p}_j$ is weak, which
    makes reaching consistency across clusters difficult. More detailed discussions on selecting $\rho$ can be found in \cite{hong2017linear}.
\end{Remark 2}

Directly following the statements in Theorem \ref{Theorem:GS-ADMM Convergence}, we can obtain the
following result on the convergence speed:
 \begin{Theorem 2}\label{Theorem:GS-ADMM Convergence Rate}
The convergence rate of GS-ADMM is $O(1/t)$, where $t$ is the
iteration time.
\end{Theorem 2}

{\it Proof}: The result can be obtained directly from the proof
of  Theorem \ref{Theorem:GS-ADMM Convergence} and is omitted.\hfill $\blacksquare$

\subsection{Convergence Analysis of J-ADMM}
To  analyze the convergence of J-ADMM, we first define several terms:
Let $\bm{p}^k=[\bm{p}_1^{kT},\bm{p}_2^{kT},...,\bm{p}_m^{kT} ]^T$ and
$\bm{\lambda}^k=[\bm{\lambda}_{i,j}^k]_{ij,e_{i,j}\in E}$ be the
results for \eqref{eq:J-ADMM pi update} and \eqref{eq:J-ADMM lambda update}
 for iteration $k$. Augment the coefficients $\gamma_i$ of proximal terms into a matrix $Q_P={\rm diag}\{\gamma_1I_D, \gamma_2I_D, ...,\gamma_mI_D \}$
 and introduce a positive definite diagonal matrix $Q_C={\rm diag}\{|B_1|I_D, |B_2|I_D, ...,|B_m|I_D \}$,
 where $|B_i|$ is the number of clusters in $B_i$. Since $Q_C$ and $Q_P$ are both diagonal matrices, we can define
 a new diagonal matrix $\bar Q$ according to $\bar Q^T\bar Q=Q_C+I+Q_P$ where $I$ is the identity matrix. It can be easily verified that $\bar Q$
 has the following form:
\begin{equation}\label{eqn:bar_Q}
\bar Q=\textrm{diag}\{\gamma_1'I_D, \gamma_2'I_D, ...,\gamma_m'I_D\},
\end{equation}
 with $\gamma_i'>0$ for $i=1,2,\ldots,m$.
  Assuming that the original problem \eqref{eq:distributed admm form} admits a solution $(\bm{p}^*,\bm{\lambda}^*)$, then we have the following
  theorem:

 \begin{Theorem 3}\label{Theorem:J-ADMM Convergence}  Let
$\bar{\bm{p}}^{t+1}=\frac{1}{t+1}\sum\limits_{k=0}^{t}\bm{p}^{k+1}$
be the  average of $\bm{p}^{k}$ up to iteration time $t+1$ and
denote the  eigenvalues of $C^TC$ as $\alpha_i$. If $\gamma_i'\ge \sqrt{\alpha_{\max}}$ is true with $\alpha_{\max}=\max\{\alpha_i\} $,
then the following holds for all $t$:
\begin{eqnarray}
 0\le L(\bar{\bm{p}}^{t+1},\bm{\lambda}^*)-L(\bm{p}^*,\bm{\lambda}^*)\le \frac{c_1}{t+1},
\end{eqnarray}
where  $L(\bm{p},\bm{\lambda})=f(\bm{p})+\bm{\lambda}^TCJ\bm{p}$ is
the Lagrangian function, and
\begin{eqnarray}
 c_1=\frac{1}{2\rho}\parallel \bm{\lambda}^0-\bm{\lambda}^*\parallel^2+\frac{\rho}{2}(\parallel \bar{Q}J(\bm{p}^0-\bm{p}^*)\parallel^2 .\label{eq:c1} 
\end{eqnarray}
\end{Theorem 3}

{\it Proof}: See Appendix B. \hfill $\blacksquare$

From Theorem \ref{Theorem:J-ADMM Convergence}, we can easily obtain the following results on the
convergence speed:

 \begin{Theorem 4}\label{Theorem:J-ADMM Convergence Rate} The convergence rate of J-ADMM is $O(1/t)$, where
$t$ is the iteration time.
\end{Theorem 4}

{\it Proof}: The result can be obtained directly from the proof of
Theorem \ref{Theorem:J-ADMM Convergence} and is omitted. \hfill
$\blacksquare$

Since $c_0$ and $c_1$ are of the same form, Remark \ref{Remark:2}
for GS-ADMM also applies to the J-ADMM case. Next, we use numerical
results to evaluate the performance of GS-ADMM and
J-ADMM.

\section{SIMULATION RESULTS}
\begin{figure}[!t]
\centering
		\includegraphics[width=8cm]{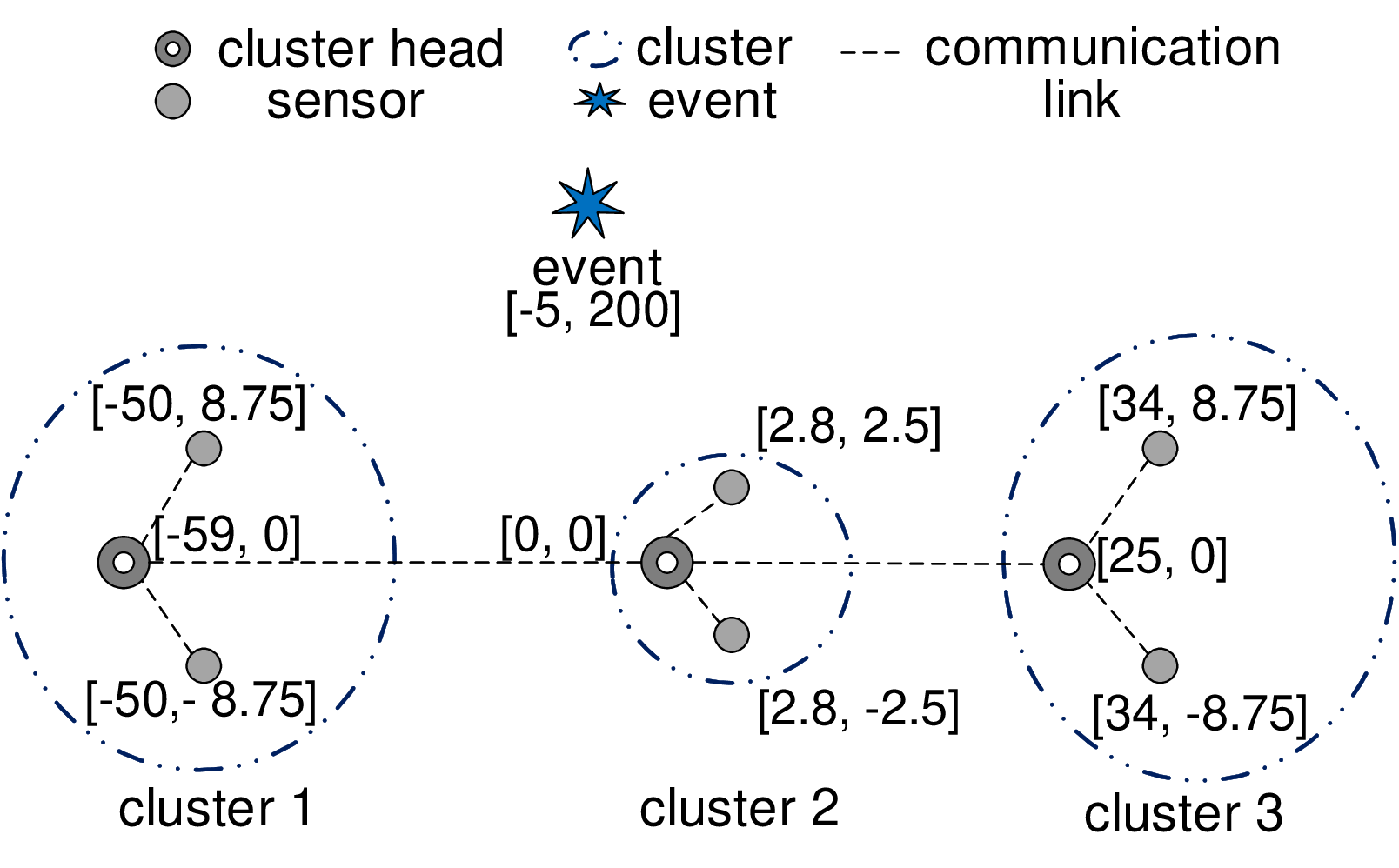}
	\caption{Event localization architecture used in simulations. The values in $[\bullet]$ denote positions (x, y coordinates) of sensors.}
	\label{fig:simulation model} \end{figure}

In this section, we illustrate effectiveness of the proposed approaches using comparison with existing results. A typical type of distributed algorithms for event localization is the projection-based algorithms. However, some projection-based algorithms, e.g., the DAPA algorithm in \cite{zhang2015distributed}, is found in our simulations not appropriate for the considered case where the target event lies outside the convex hull of sensors. More specifically, we set the sensor localization architecture similar as in \cite{NúñezF2016workingpaper,cakiades2012fusion}, which considers a practical acoustic event localization system (see Fig. \ref{fig:simulation model} for the detailed spatial distribution of all sensor nodes). The target event occurs at $\bm{x}=[-5;200]$, which is far away from the nine sensors. Simulation results suggested that DAPA did not work well in this architecture, even if we set the initial values close to the target event and used the range measurements without noise, although it did work very well
if the target event was set in the convex hull of sensors.  In the simulation, we used the same parameters for DAPA as in \cite{zhang2015distributed}, i.e., $\alpha_1=...=\alpha_9=\frac{1}{t+2}$, $\beta_1=...=\beta_9=\frac{1}{t+1}$, $b_1=...=b_9=1$, and $\xi_1=...=\xi_9=3$.

Then, we compared the localization performance of the proposed algorithms
GS-ADMM and J-ADMM  with two other projection-based algorithms: the PPM algorithm proposed in \cite{jia2011set} and the PONLM algorithm proposed in \cite{shi2008distributed}, which gave reasonable performance in the simulations.  PPM is a parallel projection method which requires a central node to average the local event location estimates obtained from all sensors in every iteration. PONLM is a sequential projection-based algorithm which solves the event localization problem by finding a point at the intersection of sensing circles. Both localization error (differences between estimated and actual target event
positions) and  localization consistency (differences in estimated
positions between clusters) are compared under different noise standard deviations $\sigma_{i,k}$.  The convergence performance is evaluated by exploring the evolution of the localization error  with  iteration time $t$.

To facilitate comparison, we first
define two performance indices:

\emph{Localization Error}: we use the root mean square error (RMSE)
to quantify the error between estimated and true positions for every cluster or sensor, which is
denoted as $\rm{ERR_{RMSE}}$:
\begin{eqnarray}
{\rm ERR_{RMSE}}  = \sqrt{\frac{\sum\limits_{j=1}^{L}\parallel
\bm{x}_{j}-\bm{x}^* \parallel^2}{L}}, \nonumber
\end{eqnarray}
where $L$ is the number of Monte Carlo trials, $\bm{x}_{j}$ is the
estimated position in the $j$th Monte Carlo trial in a certain
cluster or sensor, and $\bm{x}^*$ is the true position of the target event.

\emph{Localization Inconsistency}: We also use the root mean square
error (RMSE) to  quantify the localization inconsistency
(difference) in estimated event positions between $m$ clusters,
which is denoted as $\rm{INC_{RMSE}}$:
\begin{eqnarray}
{\rm INC_{RMSE}}  =
\sqrt{\frac{\sum\limits_{k=1}^{L}\sum\limits_{i=1}^{m-1}\sum\limits_{j=i+1}^{m}\parallel
\bm{x}_{i,k}-\bm{x}_{j,k}\parallel^2}{L}}, \nonumber
\end{eqnarray}
where $L$ is the number of Monte Carlo trials, $\bm{x}_{i,k}$ is the
estimated position obtained from the $i$th cluster in the $k$th
Monte Carlo trial. $m$ is the number of clusters.

\subsection{Convergence performance}
\begin{figure}[!h]
\centering
		\includegraphics[width=0.43\textwidth]{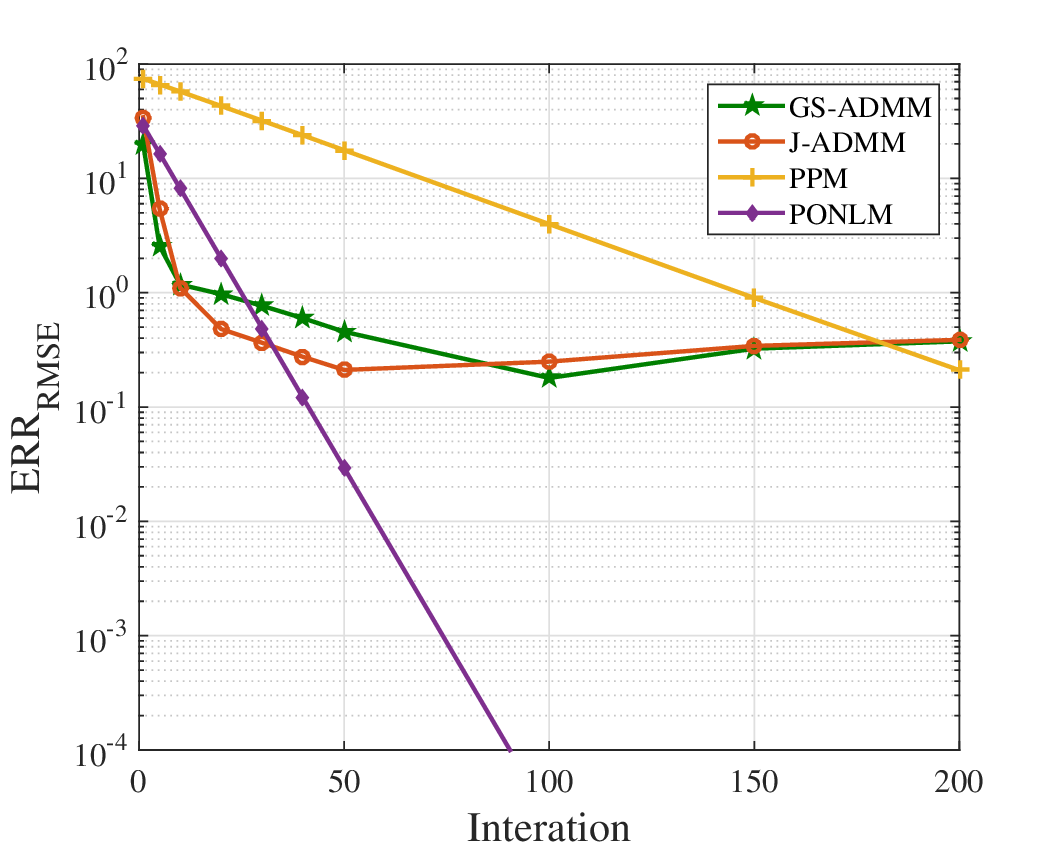}
	\caption{The evolution of localization error}
	\label{fig:ERR vs Iteration} \end{figure}

We compared the convergence performance of our sequential GS-ADMM algorithm, parallel J-ADMM algorithm, the sequential PONLM algorithm in \cite{shi2008distributed}, and the parallel PPM algorithm  in \cite{jia2011set}. For GS-ADMM and J-ADMM, we set $\rho=10^{-3}$. For PPM and PONLM, we set the initial point at $[-50; 100]$ (PPM and PONLM are sensitive to initialization settings, which will be shown later). We used the range measurements without noise in this part. The simulation results are given in Fig. \ref{fig:ERR vs Iteration}.

From Fig.  \ref{fig:ERR vs Iteration}, we can see that both GS-ADMM and J-ADMM  reached an accuracy of $10^{0}$ after about 10 iterations, while PONLM took 25 iterations and PPM took about 150 iterations. Note that sensors and clusters have to exchange local estimates in each iteration, so the required communication overhead is heavier with an increase in iteration times. The same conclusion can be drawn for energy consumption. It is worth noting that both PPM and PONLM can reach very high accuracies. However, in practical applications like gunfire localization, the accuracy of $10^0$ is sufficient \cite{cakiades2012fusion}.

\begin{Remark 2} \label{Remark:4}
	In our simulations, we used the Sedumi solver in Yalmip, whose limited precision may lead to approximate minima when solving subproblems \eqref{eq:GS-ADMM pi update} and \eqref{eq:J-ADMM pi update}. This may also lead to a low convergence speed or even fluctuations after a certain number of iterations. In addition, SeDuMi may sometimes return the message ``Run into numerical problems", which implies that
	it has terminated before it finds an approximate optimal solution \cite{waki2012generate}. In this situation, we can transform semi-definite inequality constraints into definite inequality constraints by introducing a constant positive definite term (e.g., $10^{-6}$) as indicated in \cite{labit2002user}. However, such a transformation may bring fluctuations to the convergence process.
\end{Remark 2}

\subsection{The influence of noise level on $\rm{ERR_{RMSE}}$}
\begin{figure}[!h]
	\centering
	\includegraphics[width=7cm]{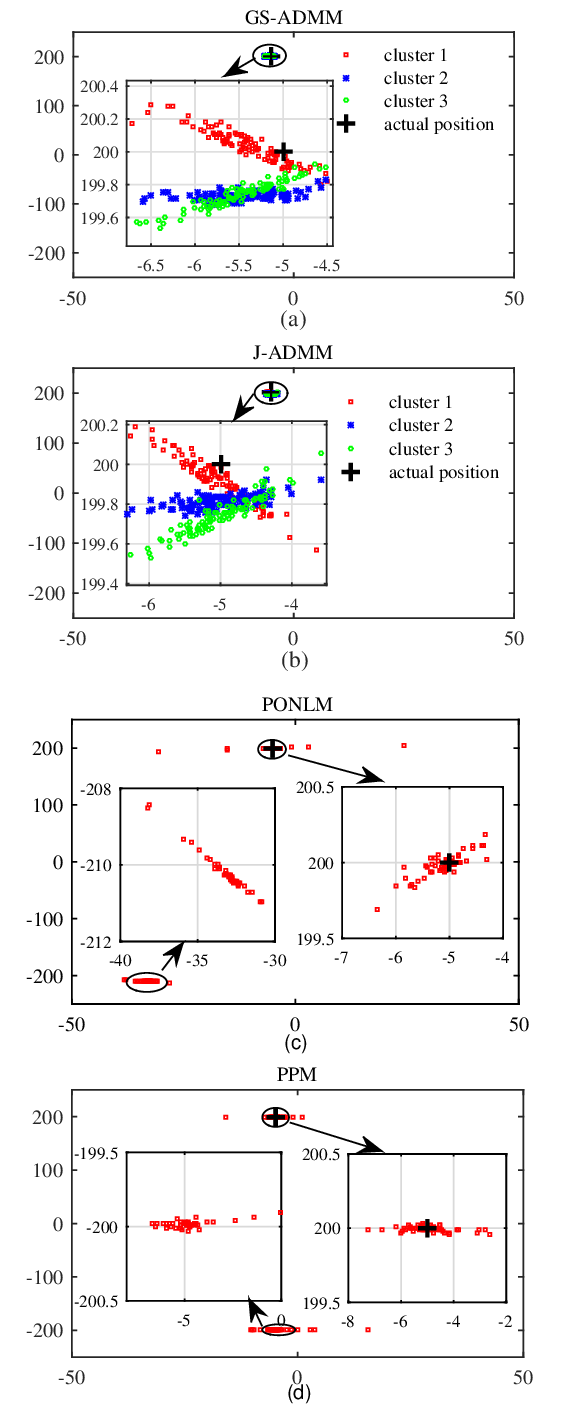}
	\caption{The distribution of estimated event location, $\sigma=0.05$. (a) GS-ADMM; (b) J-ADMM; (c) PONLM; (d) PPM.}
	\label{fig:distribution of estimated event location} \end{figure}

In this section, we simulated the event localization
algorithms under different levels of Guassian noise standard
deviation ${\sigma}_{i,k}$. For GS-ADMM and J-ADMM, we set $\rho=10^{-3}$. For PPM and PONLM, we ran simulations under two cases: setting fixed initial values at $[-50;100]$ (denote as Fix in Table \ref{tab:ERR vs Noise}) and setting random initial values in the area of 10000m $\times$ 10000m (denote as Ran in Table \ref{tab:ERR vs Noise}). The number of iterations is fixed to 50 for GS-ADMM, J-ADMM, PONLM, and 200 for PPM.  All simulation results are
summarized in Table \ref{tab:ERR vs Noise} and Fig. \ref{fig:distribution of estimated event location}. Each data point in Table \ref{tab:ERR vs Noise} is an average of 100  Monte Carlo trials.

\begin{table*}[h]\small
	\centering
	\caption{$\rm{ERR_{RMSE}}$ of GS-ADMM, J-ADMM, PPM, and PONLM under different measurement noise}
	\label{tab:ERR vs Noise}
	\begin{tabular}{|c|c|c|c|c|}
		\hline
		${\sigma}_{i,k}$ &\textbf{GS-ADMM} &\textbf{J-ADMM} & \textbf{PPM} & \textbf{PONLM }\\
		\hline
		\quad& $   \rm{CL}_{1}   \quad\quad \rm{CL}_{2}  \quad\quad  \rm{CL}_{3} $ &$  \rm{CL}_{1}   \quad\quad \rm{CL}_{2}  \quad\quad  \rm{CL}_{3} $ & \text{Fix} \quad\quad \text{Ran} & \text{Fix} \quad\quad \text{Ran}\\
		\hline
		0.00& 0.3693 \quad0.4848 \quad0.5128&  0.1146\quad  0.2223\quad 0.2964  & 0.2100 \quad273.19& 0.0338\quad 304.52\\
		\hline
		0.01 & 0.5417 \quad0.5443\quad 0.5753& 0.3546\quad 0.3911 \quad0.4385 & 0.2106\quad 285.65& 0.0498\quad 269.86\\
		\hline
		0.02&  0.5453\quad 0.5862\quad 0.5992&  0.2766\quad 0.3266 \quad0.3779 & 0.2145\quad 282.83& 0.0865\quad 307.95\\
		\hline
		0.05& 0.6055\quad 0.6723 \quad0.7188& 0.4987\quad 0.5261 \quad0.5724 & 0.2265 \quad268.31 & 0.1895 \quad278.84 \\
		\hline
		0.10&  1.0564\quad 1.0942\quad 1.1440&  1.0562 \quad1.0589\quad 1.1017 & 0.2832 \quad288.41 & 0.4019\quad 243.67\\
		\hline
	\end{tabular}
\end{table*}

%

From Table \ref{tab:ERR vs Noise},  we can see that both
PPM and PONLM  reached high localization accuracies under fixed initial values. However, their performance deteriorated significantly when random initial values were used. Therefore PPM and PONLM are sensitive to initial value settings. If the target event lies outside the convex hull of sensors, the convergent values of PPM and PONLM may be far away from the true event position. GS-ADMM and J-ADMM can avoid the convex hull problem, so every estimate lay close to the true event position.

Fig. \ref{fig:distribution of estimated event location} visualizes the estimated event locations. Fig. \ref{fig:distribution of estimated event location} (a) and (b) show the localization results  of the proposed algorithms GS-ADMM and J-ADMM respectively from 100 Monte Carlo trials with $\rho=10^{-3}$. Fig. \ref{fig:distribution of estimated event location} (c) and (d) show the results of PPM and PONLM respectively where the initial positions are chosen  randomly. 
It is clear that both GS-ADMM and J-ADMM performed better than PPM and PONLM when the initial values are randomly chosen.

\subsection{The influence of noise level on $\rm{INC_{RMSE}}$}
\begin{figure}[!h]
	\begin{center}
		\includegraphics[width=0.43\textwidth]{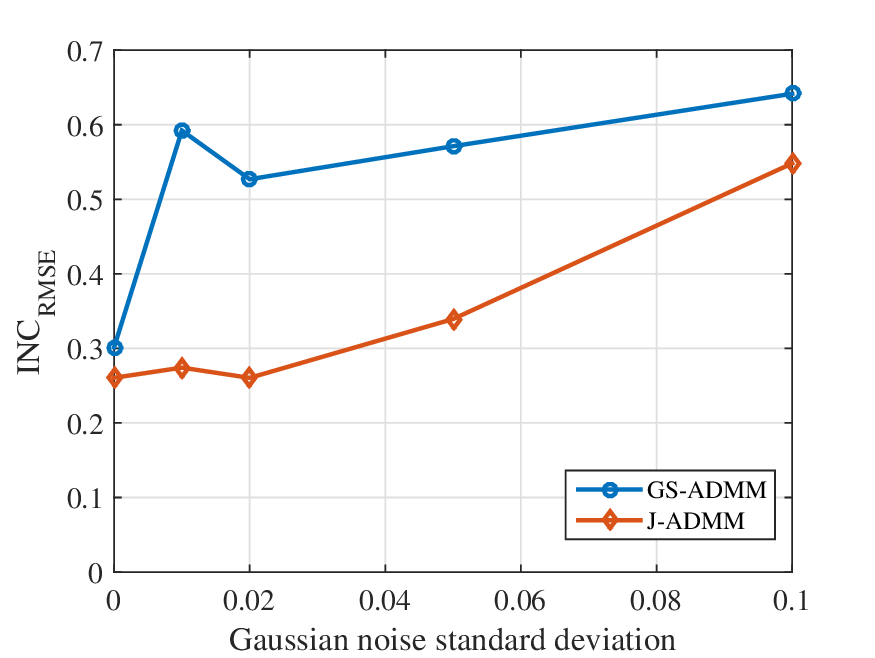}
	\end{center}
	\caption{The influence of measurement noise on localization inconsistency}
	\label{fig:INC vs noise} \end{figure}

Setting $\rho=10^{-3}$,  we also evaluated the influence of noise
level on $\rm{INC_{RMSE}}$ of our proposed algorithms. The results are summarized in   Fig. \ref{fig:INC vs noise}.

Fig. \ref{fig:INC vs noise}  indicates
that the proposed GS-ADMM and J-ADMM have small localization
inconsistency ($\rm{INC_{RMSE}}$) under different
noise strength. In other words,  our proposed algorithms
GS-ADMM and J-ADMM can achieve good consistency across clusters even
under large noise standard deviations. As indicated before, consistency is of
crucial importance in many applications.

\section{CONCLUSIONS}
We proposed two ADMM based distributed event localization algorithms GS-ADMM and J-ADMM that do not require the target event to be within the convex hull of the deployed sensors. Convergence properties of the algorithms are
analyzed theoretically. Numerical simulations showed that the proposed
algorithms are robust to measurement noises and insensitive to convex hull problem compared with existing projection-based algorithms.




\ifCLASSOPTIONcompsoc
\section*{Acknowledgments}
\else
\section*{Acknowledgment}
\fi

We would like to thanks  Dr. Andrea Simonetto for providing Matlab
codes for his paper \cite{simonetto2014distributed}.

\appendices
\section{Proof of \eqref{eq:item 1} in Theorem \ref{Theorem:GS-ADMM Convergence}}
To prove \eqref{eq:item 1} in Theorem  \ref{Theorem:GS-ADMM
	Convergence}, we first introduce two lemmas:
\begin{Lemma 1}\label{Lemma_1}
		Let $\bm{p}^k=[\bm{p}_1^{kT},\bm{p}_2^{kT},...,\bm{p}_m^{kT} ]^T$ and $\bm{\lambda}^k=[\bm{\lambda}_{i,j}^k]_{ij,e_{i,j}\in E}$ be
		the iterates generated by GS-ADMM following \eqref{eq:GS-ADMM pi update} and \eqref{eq:GS-ADMM lambda update}, then
		the following inequality holds for all $k$:
		\begin{equation} 
		\begin{aligned}
		& f(\bm{p})-f(\bm{p}^{k+1})+(\bm{p}-\bm{p}^{k+1})^TJ^TC^T\bm{\lambda}^{k+1}+ \label{eq:lemma 1} \\
		&\rho(\bm{p}-\bm{p}^{k+1})^TJ^T(-C^TH+H^TH+I)J(\bm{p}^{k+1}-\bm{p}^{k})\ge 0 ,\\
		& \forall \bm{p}\in\{[\bm{p}_1^{T},\bm{p}_2^{T},...,\bm{p}_m^{T} ]^T|\bm{p}_i\in\mathcal{P}_i,\forall i\in\{1,2,...,m\}\},
		\end{aligned}
		\end{equation}
		where $C$ is the edge-node incident matrix defined in \eqref{eq:C}, $H=\min\{0,C\}$, and $I$ is the identity matrix.
		(In the following, we only consider $\bm{p}$ belonging to the set $\{[\bm{p}_1^{T},\bm{p}_2^{T},...,\bm{p}_m^{T} ]^T|\bm{p}_i\in\mathcal{P}_i,\forall i\in\{1,2,...,m\}\}$, so we leave out this constraint in the following lemmas and proofs.)
\end{Lemma 1}

{\it Proof}: Denote by $g_i$ the function
\begin{equation} 
\begin{aligned}
&  g_i^k(\bm{p}_i) =\sum\limits_{j\in{\hat{B}_i,j\ge i}}(\bm{\lambda}_{i,j}^{kT}(J_i\bm{p}_i-J_j\bm{p}_{j}^k)+ \frac{\rho}{2}\parallel J_i\bm{p}_i-J_j\bm{p}_{j}^k \parallel ^2) \label{eq:gi GS-ADMM} \\
& \quad +\sum\limits_{j\in{\hat{B}_i,j<i}}(\bm{\lambda}_{i,j}^{kT}(J_i\bm{p}_i-J_j\bm{p}_{j}^{k+1})+ \frac{\rho}{2}\parallel J_i\bm{p}_i-J_j\bm{p}_{j}^{k+1} \parallel ^2) . 
\end{aligned}
\end{equation}

From the update rule in \eqref{eq:GS-ADMM pi update}, we know that
$\bm{p}_i^{k+1}$ is the optimizer of $g_i^k+f_i$ in the closed and convex set $\mathcal{P}_i$. Since $f_i$ and $g_i^k$ are convex, and $g_i^k$ is differentiable, following the proof of Lemma 3.1 in \cite{he20121} (which is also mentioned in Lemma 1 in \cite{mota2011proof}), we can get
\begin{equation}
f_i(\bm{p}_i)- f_i(\bm{p}_i^{k+1})+(\bm{p}_i-\bm{p}_i^{k+1})^T\bigtriangledown g_i(\bm{p}_i^{k+1})\ge 0. \nonumber
\end{equation}

Substituting $\bigtriangledown g_i(\bm{p}_i^{k+1})$ with
\eqref{eq:gi GS-ADMM}, we have
\begin{equation}
\begin{aligned}
& f_i(\bm{p}_i)- f_i(\bm{p}_i^{k+1})+(\bm{p}_i-\bm{p}_i^{k+1})^T\bm{\cdot}  \nonumber\\
&(\sum\limits_{j\in{\hat{B}_i},j\ge i}(J_i^T\bm{\lambda}_{i,j}^k+ \rho J_i^T( J_i\bm{p}_i^{k+1}-J_j\bm{p}_{j}^k )))+(\bm{p}_i-\bm{p}_i^{k+1})^T\bm{\cdot} \nonumber \\
& (\sum\limits_{j\in{\hat{B}_i},j<i}(J_i^T\bm{\lambda}_{i,j}^{k}+ \rho J_i^T( J_i\bm{p}_i^{k+1}-J_j\bm{p}_{j}^{k+1})))\ge 0. \nonumber
\end{aligned}
\end{equation}

Noting $\bm{\lambda}_{i,i}=0$, using \eqref{eq:GS-ADMM lambda update}
leads to
\begin{eqnarray}
\lefteqn{ f_i(\bm{p}_i)- f_i(\bm{p}_i^{k+1})+(\bm{p}_i-\bm{p}_i^{k+1})^T\bm{\cdot}} \nonumber\\
&& (\sum\limits_{j\in{B_i}}J_i^T\bm{\lambda}_{i,j}^{k+1}+\sum\limits_{j\in{\hat{B}_i},j\ge i} \rho J_i^T( J_j\bm{p}_j^{k+1}-J_j\bm{p}_{j}^k ))\ge 0. \nonumber
\end{eqnarray}

Noting $\bm{\lambda}_{i,j}=-\bm{\lambda}_{j,i}$, from the
definition of $C$, we can rewrite the above inequality as
\begin{equation} 
\begin{aligned}
\lefteqn{ f_i(\bm{p}_i)- f_i(\bm{p}_i^{k+1})+(\bm{p}_i-\bm{p}_i^{k+1})^T\bm{\cdot} } \label{eq:1} \\
& \quad (J_i^T[C]_i^T\bm{\lambda}^{k+1}+\sum\limits_{j\in{\hat{B}_i},j\ge i} \rho J_i^T( J_j\bm{p}_j^{k+1}-J_j\bm{p}_{j}^k ))\ge 0, 
\end{aligned}
\end{equation}
here  $[C]_i$  denotes the columns of $C$ associated with cluster $i$.

Summing both sides of 
\eqref{eq:1} over $i=1,2,...,m$, and noticing that the following two equations hold \cite{wei2012distributed},
\begin{eqnarray}
\lefteqn{ \sum\limits_{i=1}^{m}(\bm{p}_i-\bm{p}_i^{k+1})^TJ_i^T[C]_i^T\bm{\lambda}^{k+1}=(J\bm{p}-J\bm{p}^{k+1})^TC^T\bm{\lambda}^{k+1}, }\nonumber\\
\lefteqn{ \sum\limits_{i=1}^{m}(\bm{p}_i-\bm{p}_i^{k+1})^T(\sum\limits_{j\in{\hat{B}_i},j\ge
		i} \rho J_i^T( J_j\bm{p}_j^{k+1}-J_j\bm{p}_{j}^k ))} \nonumber \\
&&=\rho(J\bm{p}-J\bm{p}^{k+1})^T[(-C+H)^TH+I](J\bm{p}^{k+1}-J\bm{p}^{k}), \nonumber
\end{eqnarray}
we can get the lemma. \hfill $\blacksquare$

\begin{Lemma 2}\label{Lemma_2}
	Let $\bm{p}^k=[\bm{p}_1^{kT},\bm{p}_2^{kT},...,\bm{p}_m^{kT} ]^T$  and $\bm{\lambda}^k=[\bm{\lambda}_{i,j}^k]_{ij,e_{i,j}\in E}$ be the iterates generated by GS-ADMM following \eqref{eq:GS-ADMM pi update} and \eqref{eq:GS-ADMM lambda update}, then the following equality holds for all
	$k$:
	\begin{equation}\label{eq:lemma 2}
	\begin{aligned}
	&-(J\bm{p}^{k+1})^TC^T(\bm{\lambda}^{k+1}-\bm{\lambda}^{*})   \\
	& +\rho(J\bm{p}^{*}-J\bm{p}^{k+1})^T(H^TH-C^TH+I)J(\bm{p}^{k+1}-\bm{p}^{k})  \\
	&=-\frac{1}{2\rho}(\parallel \bm{\lambda}^{k+1}-\bm{\lambda}^*\parallel^2-\parallel \bm{\lambda}^{k}-\bm{\lambda}^*\parallel^2) \\
	&-\frac{\rho}{2}(\parallel HJ(\bm{p}^{k+1}-\bm{p}^*)\parallel^2-\parallel HJ(\bm{p}^{k}-\bm{p}^*)\parallel^2) \\
	& -\frac{\rho}{2}(\parallel J\bm{p}^{k+1}-J\bm{p}^*\parallel^2-\parallel J\bm{p}^{k}-J\bm{p}^*\parallel^2)  \\
	&-\frac{\rho}{2}\parallel HJ(\bm{p}^{k+1}-\bm{p}^k)-CJ\bm{p}^{k+1}\parallel^2 \\
	&-\frac{\rho}{2}\parallel J\bm{p}^{k+1}-J\bm{p}^k \parallel^2 .
	\end{aligned}
	\end{equation}
\end{Lemma 2}

{\it Proof}:  Since for a scalar $a$, $a^T=a$ holds, and recall
$\bm{\lambda}^{k+1}=\bm{\lambda}^k+\rho CJ\bm{p}^{k+1}$, we can get
\begin{equation}
\begin{aligned}
\lefteqn{ (\bm{p}^{k+1})^TJ^TC^T(\bm{\lambda}^{k+1}-\bm{\lambda}^{*})}\\
& \qquad
=\frac{1}{\rho}(\bm{\lambda}^{k+1}-\bm{\lambda}^k)^T(\bm{\lambda}^{k+1}-\bm{\lambda}^{*}). \label{eq:lemma2 1}
\end{aligned}
\end{equation}

In addition, as $(\bm{p}^*,\bm{\lambda}^*)$ is the saddle point of the Lagrangian function $L(\bm{p},\bm{\lambda})=f(\bm{p})+\bm{\lambda}^TCJ\bm{p}$, we have
$CJ\bm{p}^*=0$. So we can establish the following
relationships using algebraic manipulation:
\begin{equation}
\begin{aligned}
\lefteqn{(\bm{\lambda}^{k+1}-\bm{\lambda}^{k})^T(\bm{\lambda}^{k+1}-\bm{\lambda}^{*})=\frac{1}{2}\parallel \bm{\lambda}^{k+1}-\bm{\lambda}^{k}\parallel^2} \qquad\\
&\qquad+\frac{1}{2}(\parallel \bm{\lambda}^{k+1}-\bm{\lambda}^*\parallel^2-\parallel \bm{\lambda}^{k}-\bm{\lambda}^*\parallel^2), \label{eq:lemma2 2}
\end{aligned}
\end{equation}
\begin{equation}
\begin{aligned}
\lefteqn{(\bm{p}^{k+1}-\bm{p}^{*})^TJ^TIJ(\bm{p}^{k+1}-\bm{p}^{k})=\frac{1}{2}\parallel J\bm{p}^{k+1}-J\bm{p}^{k}\parallel^2}  \\
&\qquad+\frac{1}{2}(\parallel J\bm{p}^{k+1}-J\bm{p}^*\parallel^2-\parallel J\bm{p}^{k}-J\bm{p}^*\parallel^2),\label{eq:lemma2 3} 
\end{aligned}
\end{equation}
\begin{equation}
\begin{aligned}
\lefteqn{(\bm{p}^{k+1}-\bm{p}^{*})^TJ^TH^THJ(\bm{p}^{k+1}-\bm{p}^{k})} \\
&=\frac{1}{2}(\parallel HJ( \bm{p}^{k+1}-\bm{p}^*)\parallel^2-\parallel HJ(\bm{p}^{k}-\bm{p}^*)\parallel^2) \\
&+\frac{1}{2}\parallel HJ(\bm{p}^{k+1}-\bm{p}^{k})\parallel^2, \label{eq:lemma2 4} 
\end{aligned}
\end{equation}
\begin{equation}
\begin{aligned}
\lefteqn{(\bm{p}^{k+1}-\bm{p}^{*})^TJ^TC^THJ(\bm{p}^{k+1}-\bm{p}^{k})}  \\
&=\frac{1}{2}\parallel HJ(\bm{p}^{k+1}-\bm{p}^{k})\parallel^2+\frac{1}{2\rho^2}\parallel \bm{\lambda}^{k+1}-\bm{\lambda}^{k}\parallel^2 \\
&-\frac{1}{2}\parallel HJ(\bm{p}^{k+1}-\bm{p}^k)-CJ\bm{p}^{k+1}\parallel^2 . \label{eq:lemma2 5}
\end{aligned}
\end{equation}

Then \eqref{eq:lemma 2} can be proven by plugging equations \eqref{eq:lemma2 1} to \eqref{eq:lemma2 5} 
into  the left part of \eqref{eq:lemma 2}. \hfill $\blacksquare$

Now we proceed to prove Theorem \ref{Theorem:GS-ADMM Convergence}.
Set $\bm{p}=\bm{p}^*$ in \eqref{eq:lemma 1}, and recall $CJ\bm{p}^*=0$, then we have
\begin{equation}
\begin{aligned}
\lefteqn{ f(\bm{p}^*)-f(\bm{p}^{k+1})-\bm{p}^{(k+1)T}J^TC^T\bm{\lambda}^{k+1} +} \label{eq:p=p*}\\
& \rho(\bm{p}^*-\bm{p}^{k+1})^TJ^T(-C^TH+H^TH+I)J(\bm{p}^{k+1}-\bm{p}^{k})\ge 0.
\end{aligned}
\end{equation}

Adding and subtracting the term $\bm{\lambda}^{*T}CJ\bm{p}^{k+1}$ from the left side of \eqref{eq:p=p*}, we can get
\begin{equation}
\begin{aligned}
\lefteqn{f(\bm{p}^*)-f(\bm{p}^{k+1})} \\
&	-\bm{\lambda}^{*T}CJ\bm{p}^{k+1}-\bm{p}^{(k+1)T}J^TC^T(\bm{\lambda}^{k+1}-\bm{\lambda}^*)+\nonumber\\ 
& \rho(\bm{p}^*-\bm{p}^{k+1})^TJ^T(-C^TH+H^TH+I)J(\bm{p}^{k+1}-\bm{p}^{k})\ge 0. \nonumber
\end{aligned}
\end{equation}

Now by applying \eqref{eq:lemma 2} into the above inequality, the following inequality can be obtained:
\begin{eqnarray}
&& f(\bm{p}^*)-f(\bm{p}^{k+1})-\bm{\lambda}^{*T}CJ\bm{p}^{k+1} \nonumber\\
&& -\frac{1}{2\rho}(\parallel \bm{\lambda}^{k+1}-\bm{\lambda}^*\parallel^2-\parallel \bm{\lambda}^{k}-\bm{\lambda}^*\parallel^2)  \nonumber \\
&& -\frac{\rho}{2}(\parallel HJ(\bm{p}^{k+1}-\bm{p}^*)\parallel^2-\parallel HJ(\bm{p}^{k}-\bm{p}^*)\parallel^2) \nonumber \\
&& -\frac{\rho}{2}(\parallel J\bm{p}^{k+1}-J\bm{p}^*\parallel^2-\parallel J\bm{p}^{k}-J\bm{p}^*\parallel^2)  \nonumber \\
&& -\frac{\rho}{2}\parallel HJ(\bm{p}^{k+1}-\bm{p}^k)-CJ\bm{p}^{k+1}\parallel^2 \nonumber \\
&&-\frac{\rho}{2}\parallel J\bm{p}^{k+1}-J\bm{p}^k \parallel^2 \ge 0. \nonumber
\end{eqnarray}

Summing both sides of the inequality over $k=0,1,...,t$, we can obtain the following result after some re-arrangement:
 \begin{equation} \nonumber
 \begin{aligned}
 &(t+1)f(\bm{p}^*)-\sum\limits_{k=0}^{t}f(\bm{p}^{k+1})-\bm{\lambda}^{*T}CJ\sum\limits_{k=0}^{t}\bm{p}^{k+1}+\frac{\rho}{2}\bm{\cdot}\nonumber \\
 & (\parallel HJ(\bm{p}^{0}-\bm{p}^*)\parallel^2+\parallel J\bm{p}^{0}-J\bm{p}^*\parallel^2)+\frac{1}{2\rho}\parallel \bm{\lambda}^{0}-\bm{\lambda}^*\parallel^2 \nonumber \\
 &\ge \sum\limits_{k=0}^{t}\frac{\rho}{2}(\parallel HJ(\bm{p}^{k+1}-\bm{p}^k)-CJ\bm{p}^{k+1}\parallel^2) \\
 &+\sum\limits_{k=0}^{t}\frac{\rho}{2}(\parallel J\bm{p}^{k+1}-J\bm{p}^k \parallel^2)+\frac{1}{2\rho}\parallel \bm{\lambda}^{t+1}-\bm{\lambda}^*\parallel^2\nonumber \\
 &+\frac{\rho}{2}(\parallel HJ(\bm{p}^{t+1}-\bm{p}^*)+\parallel J\bm{p}^{t+1}-J\bm{p}^*\parallel^2)\ge 0.  \nonumber
 \end{aligned}
 \end{equation}

In addition, as our function is convex, we have
$\sum\limits_{k=0}^{t}f(\bm{p}^{k+1})\ge (t+1)f(\bar{\bm{p}}^{t+1})
$, then we can get
\begin{equation}
\begin{aligned}
&(t+1)f(\bm{p}^*)-(t+1)f(\bar{\bm{p}}^{t+1})-(t+1)\bm{\lambda}^{*T}CJ\bar{\bm{p}}^{t+1}\nonumber\\
&+\frac{\rho}{2}(\parallel HJ(\bm{p}^{0}-\bm{p}^*)\parallel^2+\parallel J\bm{p}^{0}-J\bm{p}^*\parallel^2) \\
&+\frac{1}{2\rho}\parallel \bm{\lambda}^{0}-\bm{\lambda}^*\parallel^2 \ge 0 .  \nonumber
\end{aligned}
\end{equation}

Dividing both sides by $-(t+1)$ yields
\begin{equation}
\begin{aligned}
\lefteqn{f(\bar{\bm{p}}^{t+1})+\bm{\lambda}^{*T}CJ\bar{\bm{p}}^{t+1} -f(\bm{p}^*)}\label{eq:theorem 1} \\
& \le\frac{\rho}{2(t+1)}(\parallel HJ(\bm{p}^{0}-\bm{p}^*)\parallel^2 +\parallel J\bm{p}^{0}-J\bm{p}^*\parallel^2) \\
& +\frac{1}{(t+1)2\rho}\parallel \bm{\lambda}^{0}-\bm{\lambda}^*\parallel^2. 
\end{aligned}
\end{equation}

Combining the above relationship \eqref{eq:theorem 1} with the Lagrangian
function $L(\bm{p},\bm{\lambda})=f(\bm{p})+\bm{\lambda}^TCJ\bm{p}$, \eqref{eq:item 1} in Theorem \ref{Theorem:GS-ADMM Convergence} is proven.

\hfill $\blacksquare$

\section{Proof of Theorem \ref{Theorem:J-ADMM Convergence}}
To prove Theorem \ref{Theorem:J-ADMM Convergence}, we first introduce two lemmas:
\begin{Lemma 3} \label{Lemma_3}
	Let $\bm{p}^k=[\bm{p}_1^{kT},\bm{p}_2^{kT},...,\bm{p}_m^{kT} ]^T$ and $\bm{\lambda}^k=[\bm{\lambda}_{i,j}^k]_{ij,e_{i,j}\in E}$ be the iterates generated by J-ADMM  following \eqref{eq:J-ADMM pi update} and \eqref{eq:J-ADMM lambda update}, then the following inequality holds for all $k$:
	\begin{equation}
	\begin{aligned}
	\lefteqn{ f(\bm{p})-f(\bm{p}^{k+1})+(\bm{p}-\bm{p}^{k+1})^TJ^TC^T\bm{\lambda}^{k+1}} \label{eq:lemma 3} \\
	&  +\rho(\bm{p}-\bm{p}^{k+1})^TJ^T(-C^TC+\bar{Q}^T\bar{Q})J(\bm{p}^{k+1}-\bm{p}^{k})\ge 0,
	\end{aligned}
	\end{equation}
	where  $\bar{Q}$  is defined in (\ref{eqn:bar_Q}).
\end{Lemma 3}

{\it Proof}: Denote by $g_i$ the function
\begin{equation}
\begin{aligned}
\lefteqn{g_i^k(\bm{p}_i)=\sum\limits_{j\in\hat{B}_{i}}(\bm{\lambda}_{i,j}^{kT}(J_i\bm{p}_{i}-J_j\bm{p}_j^{k})} \label{eq:gi J-ADMM} \\
&\qquad+ \frac{\rho}{2}\parallel J_i\bm{p}_{i}-J_j\bm{p}_{j}^{k} \parallel ^2)+\frac{\rho\gamma_i}{2}\parallel J_i\bm{p}_i-J_i\bm{p}_i^k \parallel^2.
\end{aligned}
\end{equation}

	Then following the proof of Lemma 1, we can get
	\begin{equation} 
	\begin{aligned}
	\lefteqn{ f_i(\bm{p}_i)- f_i(\bm{p}_i^{k+1})+(\bm{p}_i-\bm{p}_i^{k+1})^TJ_i^T\bm{\cdot}} \label{eq:2} \\
	&  ([C]_i^T\bm{\lambda}^{k+1} +\sum\limits_{j\in{\hat{B}_i}} \rho J_j( \bm{p}_j^{k+1}-\bm{p}_{j}^k )+\rho \gamma_iJ_i(\bm{p}_i^{k+1}-\bm{p}_i^k))\ge 0 .
	\end{aligned}
	\end{equation}

	Summing both sides of the above relation over $i=1,2,...m$, and noticing that the following two equations hold, 
	\begin{equation}
	\begin{aligned}
	&\sum\limits_{i=1}^{m}(\bm{p}_i-\bm{p}_i^{k+1})^TJ_i^T
	\rho(\sum\limits_{j\in{\hat{B}_i}} J_j( \bm{p}_j^{k+1}-\bm{p}_{j}^k )+ \gamma_iJ_i(\bm{p}_i^{k+1}-\bm{p}_i^{k}))\nonumber \\
	&=\rho(\bm{p}-\bm{p}^{k+1})^TJ^T[-C^TC+Q_C+I+Q_P]J(\bm{p}^{k+1}-\bm{p}^{k}), \nonumber \\
	\lefteqn{\sum\limits_{i=1}^{m}(\bm{p}_i-\bm{p}_i^{k+1})^TJ_i^T[C]_i^T\bm{\lambda}^{k+1}=(\bm{p}-\bm{p}^{k+1})^TJ^TC^T\bm{\lambda}^{k+1},} \nonumber
	\end{aligned}
	\end{equation}
	we can get the lemma.\hfill $\blacksquare$

\begin{Lemma 4} \label{Lemma_4}
		Let $\bm{p}^k=[\bm{p}_1^{kT},\bm{p}_2^{kT},...,\bm{p}_m^{kT} ]^T$ and
		$\bm{\lambda}^k=[\bm{\lambda}_{i,j}^k]_{ij,e_{i,j}\in E}$ be
		the iterates generated by J-ADMM following \eqref{eq:J-ADMM pi update}
		and \eqref{eq:J-ADMM lambda update}. Then the following equality holds for all
		$k$:
		\begin{equation} 
		\begin{aligned}
		&-(\bm{p}^{k+1})^TJ^TC^T(\bm{\lambda}^{k+1}-\bm{\lambda}^{*}) \label{eq:lemma 4} \\
		& +\rho(\bm{p}^{*}-\bm{p}^{k+1})^TJ^T(-C^TC+\bar{Q}^T\bar{Q})J(\bm{p}^{k+1}-\bm{p}^{k})  \\
		&=-\frac{1}{2\rho}(\parallel \bm{\lambda}^{k+1}-\bm{\lambda}^*\parallel^2-\parallel \bm{\lambda}^{k}-\bm{\lambda}^*\parallel^2)  \\
		&+\frac{\rho}{2}(\parallel CJ(\bm{p}^{k+1}-\bm{p}^*)\parallel^2-\parallel CJ(\bm{p}^{k}-\bm{p}^*)\parallel^2)  \\
		& -\frac{\rho}{2}(\parallel \bar{Q}J(\bm{p}^{k+1}-\bm{p}^*)\parallel^2-\parallel \bar{Q}J(\bm{p}^{k}-\bm{p}^*)\parallel^2) \\
		&+\frac{\rho}{2}\parallel CJ(\bm{p}^{k+1}-\bm{p}^k )\parallel^2-\frac{\rho}{2}\parallel \bar{Q}J(\bm{p}^{k+1}-\bm{p}^k )\parallel^2 \\
		&-\frac{1}{2\rho} \parallel\bm{\lambda}^{k+1}-\bm{\lambda}^k\parallel^2. 
		\end{aligned}
		\end{equation}
\end{Lemma 4}

{\it Proof}: The proof is similar to the proof of Lemma
\ref{Lemma_2} and is omitted. \hfill $\blacksquare$

Then following the proof of Theorem 1 (setting $\bm{p}=\bm{p}^*$ in \eqref{eq:lemma 3} and applying \eqref{eq:lemma 4}), we can obtain the following inequality: 
\begin{eqnarray}
&& f(\bm{p}^*)-f(\bm{p}^{k+1})-\bm{\lambda}^{*T}CJ\bm{p}^{k+1}\nonumber\\
&& -\frac{1}{2\rho}(\parallel \bm{\lambda}^{k+1}-\bm{\lambda}^*\parallel^2-\parallel \bm{\lambda}^{k}-\bm{\lambda}^*\parallel^2) \nonumber \\
&&+\frac{\rho}{2}(\parallel CJ(\bm{p}^{k+1}-\bm{p}^*)\parallel^2-\parallel CJ(\bm{p}^{k}-\bm{p}^*)\parallel^2) \nonumber \\
&& -\frac{\rho}{2}(\parallel \bar{Q}J(\bm{p}^{k+1}-\bm{p}^*)\parallel^2-\parallel \bar{Q}J(\bm{p}^{k}-\bm{p}^*)\parallel^2)\nonumber \\
&& +\frac{\rho}{2}\parallel CJ(\bm{p}^{k+1}-\bm{p}^k ) \parallel^2-\frac{\rho}{2}\parallel \bar{Q}J(\bm{p}^{k+1}-\bm{p}^k )\parallel^2 \nonumber\\
&&-\frac{1}{2\rho} \parallel\bm{\lambda}^{k+1}-\bm{\lambda}^k\parallel^2 \ge 0. \nonumber
\end{eqnarray}

Summing both sides of the above inequality over $k=0,1,...,t$, we can get the following result after some re-arrangement: 
\begin{equation} \nonumber
\begin{aligned}
& (t+1)f(\bm{p}^*)-\sum\limits_{k=0}^{t}f(\bm{p}^{k+1})-\bm{\lambda}^{*T}CJ\sum\limits_{k=0}^{t}\bm{p}^{k+1} \nonumber \\
& +\frac{\rho}{2}\parallel\bar{Q}J(\bm{p}^{0}-\bm{p}^*)\parallel^2+\frac{1}{2\rho}\parallel \bm{\lambda}^{0}-\bm{\lambda}^*\parallel^2   \nonumber \\
& \ge \frac{\rho}{2}\parallel CJ(\bm{p}^{0}-\bm{p}^*)\parallel^2+\frac{1}{2\rho}\parallel \bm{\lambda}^{t+1}-\bm{\lambda}^*\parallel^2\nonumber \\
&  +\sum\limits_{k=0}^{t}\frac{\rho}{2}(\parallel \bar{Q}J(\bm{p}^{k+1}-\bm{p}^k)\parallel^2-\parallel CJ(\bm{p}^{k+1}-\bm{p}^k) \parallel^2) \nonumber\\
& +\frac{\rho}{2}(\parallel \bar{Q}J(\bm{p}^{t+1}-\bm{p}^*)\parallel^2 -\parallel CJ(\bm{p}^{t+1}-\bm{p}^*)\parallel^2) \nonumber \\
& +\sum\limits_{k=0}^{t}\frac{1}{2\rho}\parallel \bm{\lambda}^{k+1}-\bm{\lambda}^k\parallel^2 \nonumber \\
& \ge \sum\limits_{k=0}^{t}\frac{\rho}{2}(\parallel \bar{Q}J(\bm{p}^{k+1}-\bm{p}^k)\parallel^2-\parallel C\parallel^2\parallel J\bm{p}^{k+1}-J\bm{p}^k\parallel^2) \nonumber\\
& +\frac{\rho}{2}(\parallel \bar{Q}J(\bm{p}^{t+1}-\bm{p}^*)\parallel^2 -\parallel C\parallel^2\parallel J\bm{p}^{t+1}-J\bm{p}^*\parallel^2). \nonumber
\end{aligned}
\end{equation}

Since $\parallel C \parallel^2=\alpha_{\max}$,
$\bar{Q}$ is a diagonal matrix with $\gamma_i'\ge \sqrt{\alpha_{\max}}$, we
can get that the right hand side of the above inequality is greater than
$0$, which leads to
\begin{eqnarray}
\lefteqn{ (t+1)f(\bm{p}^*)-\sum\limits_{k=0}^{t}f(\bm{p}^{k+1})-\bm{\lambda}^{*T}CJ\sum\limits_{k=0}^{t}\bm{p}^{k+1}} \nonumber \\
&&\quad\quad\quad +\frac{\rho}{2}\parallel \bar{Q}J(\bm{p}^{0}-\bm{p}^*)\parallel^2+\frac{1}{2\rho}\parallel \bm{\lambda}^{0}-\bm{\lambda}^*\parallel^2 \ge 0.   \nonumber
\end{eqnarray}

In addition, as our function is convex, we have
$\sum\limits_{k=0}^{t}f(\bm{p}^{k+1})\ge (t+1)f(\bar{\bm{p}}^{t+1})
$ and
\begin{eqnarray}
\lefteqn{(t+1)f(\bm{p}^*)-(t+1)f(\bar{\bm{p}}^{t+1})-(t+1)\bm{\lambda}^{*T}CJ\bar{\bm{p}}^{t+1} }\nonumber\\
&&\quad\quad\quad+\frac{\rho}{2}\parallel \bar{Q}J(\bm{p}^{0}-\bm{p}^*)\parallel^2+\frac{1}{2\rho}\parallel \bm{\lambda}^{0}-\bm{\lambda}^*\parallel^2  \ge 0.   \nonumber
\end{eqnarray}	

By dividing both sides by $-(t+1)$, we can obtain
\begin{equation} \nonumber
\begin{aligned}
\lefteqn{ f(\bar{\bm{p}}^{t+1})+\bm{\lambda}^{*T}CJ\bar{\bm{p}}^{t+1} -f(\bm{p}^*)} \label{eq:theorem 3}\\
& \quad\le \frac{1}{t+1}(\frac{1}{2\rho}\parallel  \bm{\lambda}^{0}-\bm{\lambda}^*\parallel^2+\frac{\rho}{2}\parallel \bar{Q}J(\bm{p}^{0}-\bm{p}^*)\parallel^2).  
\end{aligned}
\end{equation}	

Combining the above relationship with the Lagrangian function $L(\bm{p},\bm{\lambda})=f(\bm{p})+\bm{\lambda}^TCJ\bm{p}$, we can get Theorem 3. \hfill $\blacksquare$

\bibliographystyle{unsrt}
\bibliography{abbr_bibli}
\end{document}